\documentstyle{article}
\def\half{\frac{1}{2}}
\def\nel{\in \hspace{-0.14in}/}

\def\dX2{d\vec{x}\cdot d\vec{x}}
\newcommand{\be}{\begin{equation}}
\newcommand{\ee}{\end{equation}}
\newcommand{\bea}{\begin{eqnarray}}
\newcommand{\eea}{\end{eqnarray}}
\newcommand{\bt}{\begin{tabbing}}
\newcommand{\et}{\end{tabbing}}
\newcount\sectionnumber
\def\sect
{\global\equationnumber=0
\global\advance\sectionnumber by 1
\the\sectionnumber . }
\newcount\equationnumber
\def   \num
{\eqno{\global\advance\equationnumber by 1
\left(\the\sectionnumber .\the\equationnumber \right)}}
\begin{document}
\begin{center}

\textbf {{\Large Physical Acceptability of Isolated, Static, Spherically Symmetric, \\Perfect Fluid Solutions of Einstein's Equations.}}\\
\vspace{20pt}
M.S.R. Delgaty and Kayll Lake\footnote{e-mail: lake@astro.queensu.ca}\\
\textit{Department of Physics, Queen's University, \\
Kingston Ontario, Canada K7L 3N6 }\\
\end{center} 

\rule{15cm}{0.5mm}

\begin{abstract}

We ask the following question: Of the exact solutions to Einstein's equations
extant in the literature, how many could represent the field associated with 
an isolated static spherically symmetric perfect fluid source? The candidate 
solutions were subjected to the following elementary tests: i) isotropy of the 
pressure, ii) regularity at the origin, iii) positive definiteness of the energy 
density and pressure at the origin, iv) vanishing of the pressure at some finite 
radius, v) monotonic decrease of the energy density and pressure with increasing 
radius, and vi) subluminal sound speed. A total of 127 candidate solutions were found. 
Only 16 of these passed all the tests. Of these 16, only 9  have a 
sound speed which monotonically decreases with radius. The analysis was facilitated 
by use of the computer algebra system \emph{GRTensorII}. 
\end{abstract}

\rule{15cm}{0.5mm}

\pagenumbering{arabic}

\section{Introduction} \label{int}
\bigskip

There are a fair number of static, spherically symmetric exact solutions of Einstein's
equations which have been written down in closed form.  (A detailed discussion
of methods for solving Einstein's equations in this context in both curvature and isotropic
coordinate systems can be found in Kuchowicz \cite{kn:Kuch4}. For a partial 
list of solutions see Kramer \textit{et al.} \cite{kn:ESEFE}, Finch \cite{kn:Finch} and Finch and Skea \cite{kn:FS}
\footnote{The review by Finch and Skea  is closely
related to the present paper in that it considers essentially the same
problem. Almost all of the present paper was completed before we learned of
this work. The present contribution extends this work  to
verify the physical conditions for a wider class of solutions. It is a
pleasure to thank Malcolm MacCallum who first pointed out this work to us.}.
In this paper we explore the question as to how many of these solutions satisfy the most elementary
criteria for physical acceptability. Here we consider only those solutions which can
be considered isolated in the sense that the boundary (where the pressure vanishes) occurs at a finite 
radius.  This is equivalent to demanding that the solution match onto the exterior
Schwarzschild solution \cite{kn:SchwExt} as it can be shown that a necessary
and sufficient condition for matching in the present case is that the pressure equal zero
at a finite radius. Cosmological solutions, where the pressure does not vanish or vanishes only as
$r \rightarrow \infty$, are not considered \footnote{The cosmological solutions of Einstein
and de Sitter are included only as they will often be referenced in the analysis.  A few
other solutions were
found to be cosmological after analysis and are included in the tables.  Solutions which were
originally stated to be cosmological are not included.}.  In addition, solutions
giving only partial metrics or inexact metric functions are considered incomplete
as the complete spacetime cannot be examined.  Finally we only consider
solutions described by a single metric
and have excluded from our search all interior solutions derived from joining two or
more exact solutions together.  The elementary criteria for physical
acceptability that we have used are as follows:
\begin{itemize}
\item[1.]{Since all the solutions found purport to represent a perfect fluid, we start by 
verifying the isotropy of the pressure. In notation as explained in Appendix A, $G^{r}_{r} =
G^{\theta}_{\theta}, ~G^{a}_{b}$ the Einstein tensor. The symmetry of course guarantees that $G^{\theta}_{\theta}=G^{\phi}_{\phi}$.}
\item[2.]{Since the solutions must integrate from a regular origin, we check the regularity 
of the scalars polynomial in the Riemann tensor \cite{kn:LakeMus} \cite {kn:ell} \cite{kn:schmidt}. 
Appendix A contains the conditions for regularity.}
\item[3.]{We require that both the pressure ($p$) and energy density ($\rho$) be positive definite 
at the origin.  Appendix B summarizes the standard equilibrium conditions.}
\item[4.]{To be isolated we require that the pressure reduce to zero at some finite 
boundary radius $r_b > 0$.}
\item[5.]{We require that both the pressure and energy density be monotonically 
decreasing to the boundary.}
\item[6.]{We require a subluminal sound speed ($v_s^2 = \frac{dp}{d\rho}
< 1$).  Whereas at very high densities the adiabatic sound speed may not equal the
actual propagation speed of the signal \cite{kn:C-B}, we do not distinguish
between these cases in this paper.}
\end{itemize}
It was found that all solutions that were studied which satisfied criteria one through three also satisfied 
the dominant energy condition ($p/\rho < 1$ for all $r < r_b$ here).

\section{Metrics} \label{sec-revsol}
\bigskip

The following tables give the name and reference of the solution studied,
along with the corresponding metric.  We have named the solutions with the
authors name (with a number to differentiate between multiple publications
by one author), followed by a solution number.  If an author rederived a
previously known solution, we have given that solution a number, but not
included it in the tables.  (For example, Kuchowicz wrote a series of five
papers on static, spherically symmetric solutions.  The first solution in his
first paper in the series \cite{kn:Kuch} is identical to Tolman VI.  The second
solution of that paper is derived in the same manner, but is apparently a new
solution.  We named the first solution of this paper Kuch1 Ia, but did not
include it in the tables, while Kuch1 Ib is found below.) The metric
names are repeated with the references and are presented in chronological order.

Appendix C contains
a brief list of rediscoveries which were not recognized by the authors.  Rediscoveries
such as Kuch1 Ia$\equiv$Tolman VI which were properly acknowledged are \textit{not} included
in this Appendix. We make \textit{no} claim that Appendix C is comprehensive. 

All but six\footnote{ Buch3 and R-R III-VIII all of which fail the tests.} of the spacetimes that we have examined turn out to be in either isotropic or curvature
coordinates (as explained in Appendix A). The choice of coordinates can be read immediately from
the following tables: If the coefficient of $d\Omega^2$ is $r^2$ then the coordinates are curvature
coordinates. It is worth noting that our search was \textit{not} restricted by choice of coordinates.

Finally, note that the style in which some of the solutions are presented has not been
optimized, but rather for the most part reflects the forms presented in the original papers.

\newpage

\vspace{0.25in}
\begin{center}
{\noindent
\begin{tabular}{|r|l|} \hline \hline
{\bf name [ref.]} & {\bf metric} \\ \hline
Schw Int. \footnotemark \cite{kn:SchwInt} & $-\left(A-B\sqrt{1- \frac{r^2}{R^2}} \right) ^2 dt^2 + \left( 1 - \frac{r^2}{R^2} \right) ^{-1} dr^2 + r^2d\Omega ^2$ \\ \hline
Einstein \footnotemark \cite{kn:Einstein}   & $-c^2dt^2 + \left( 1-\frac{r^2}{R^2}\right) ^{-1}dr^2 + r^2d\Omega ^2$ \\ \hline
de Sitter \footnotemark \cite{kn:deSitter} & $-\left( 1-\frac{r^2}{R^2}\right)dt^2 + \left( 1-\frac{r^2}{R^2}\right) ^{-1}dr^2 + r^2d\Omega ^2$ \\ \hline
Kottler \footnotemark \cite{kn:Kottler}  & $-c^2\left( 1 - \frac{2m}{r} - \frac{r^2}{R^2}\right) dt^2 + \left( 1-\frac{2m}{r} - \frac{r^2}{R^2}\right) ^{-1}dr^2 + r^2d\Omega ^2$ \\ \hline
Tolman IV \cite{kn:Tolman}  & $-B^2\left( 1 + \frac{r^2}{A^2} \right) dt^2 + \frac{1 + 2\frac{r^2}{A^2}}{\left( 1-\frac{r^2}{R^2}\right) \left( 1 + \frac{r^2}{A^2}\right)} dr^2 + r^2d\Omega ^2$ \\ \hline
Tolman V \cite{kn:Tolman}   & $-B^2r^{2n}dt^2 + \frac{1 + 2n - n^2}{ 1 - (1+2n-n^2)\left( \frac{r}{R}\right) ^{2\frac{1+2n - n^2}{1+n}}}dr^2 + r^2d\Omega ^2$ \\ \hline
Tolman VI \cite{kn:Tolman}  & $-(Ar^{1-n} - Br^{1+n})^2dt^2 + (2-n^2)dr^2 + r^2d\Omega ^2$ \\ \hline
Tolman VII \cite{kn:Tolman} & $-B^2 \sin ^2 \ln \sqrt{\frac{\sqrt{1 - \frac{r^2}{R^2} + 4\frac{r^4}{A^4}} + 2\frac{r^2}{A^2} - \frac{1}{4}\frac{A^2}{R^2}}{C}} dt^2 +$ \\
			    & $\left( 1-\frac{r^2}{R^2} + 4\frac{r^4}{A^4}\right) ^{-1}dr^2 + r^2d\Omega ^2$ \\ \hline
Tolman VIII \cite{kn:Tolman}& $-B^2r^{2b}\left( \frac{2}{(a-b)(a+2b-1)} - \left( \frac{2m}{r}\right) ^{a + 2b -1} - \left( \frac{r}{R}\right) ^{a-b} \right) dt^2 +$ \\
			    & $\left( \frac{2}{(a-b)(a+2b-1)} - \left( \frac{2m}{r}\right) ^{a + 2b -1} - \left( \frac{r}{R} \right) ^{a-b}\right) ^{-1}dr^2 + r^2d\Omega ^2$ \\ 
			    & where $(a+b)(a-1) - 2b -2 = 0 $ \\ \hline
N-P-V Ia \cite{kn:NPV} & $-\left( ar^{1+\frac{x}{2}}+br^{1-\frac{x}{2}}\right) ^2 \left( Ar^{1+\frac{n}{2}}+Br^{1-\frac{n}{2}}\right) ^{-2} dt^2 +$ \\
$\sqrt{2} < n \leq 2$    & $\left( Ar^{1+\frac{n}{2}} + Br^{1-\frac{n}{2}}  \right) ^{-2} (dr^2 + r^2d\Omega ^2)$, \\ 
			  & where $x=\sqrt{2n^2-4}$ \\ \hline
N-P-V Ib \cite{kn:NPV} & $-(a+b\ln r)^2\left( Ar^{\frac{1}{\sqrt{2}}}+Br^{-\frac{1}{\sqrt{2}}}\right) ^{-2} dt^2 +$ \\
$n = \sqrt{2}$         & $\left( Ar^{1+\frac{1}{\sqrt{2}}} + Br^{1-\frac{1}{\sqrt{2}}}  \right) ^{-2}(dr^2 + r^2d\Omega ^2)$ \\ \hline
N-P-V Ic \cite{kn:NPV} & $-\left( ar^{\frac{n}{2}}+br^{-\frac{n}{2}}\right) ^2r^4 \left[ A\cos \left( \frac{\chi}{2}\ln r\right) + B\sin \left(\frac{\chi}{2}\ln r\right) \right] ^{2} dt^2 +$ \\
$0 < n < \sqrt{2}$     & $\left( Ar^{1+\frac{n}{2}} + Br^{1-\frac{n}{2}}  \right) ^{-2}(dr^2 + r^2d\Omega ^2)$, \\ 
		       & where $\chi = \sqrt{4-2n^2}$ \\ \hline
\end{tabular}
}
\end{center}
\addtocounter{footnote}{-3}\footnotetext{Originally given in the form $-c^2\left(\frac{3\cos ^2\chi _a-\cos ^2\chi}{2} \right) ^2 dt^2 + \frac{3}{x\rho _o}\left[ d\chi ^2 + \sin ^2\chi d\theta ^2 + \sin ^2\chi\sin ^2\theta d\phi^2 \right]$ where $\sin ^2\chi = \frac{r^2}{R^2}$. Throughout this paper we use the notation $d\Omega^2=d\theta ^2 + \sin ^2\theta d\phi^2$.}
\addtocounter{footnote}{1}\footnotetext{Originally given in the form of $g_{tt} = -1, g_{ab} = -\left( \delta_{ab} + \frac{x_{a}x_{b}}{R^2-\sum x_a^2} \right), a,b \neq t$ and is a cosmological solution with the cosmological constant not zero for positive pressure and density.}
\addtocounter{footnote}{1}\footnotetext{Originally given in the form of $g_{\mu\nu} = -\left( \delta_{\mu\nu} + \frac{x_{\mu}x_{\nu}}{R^2-\sum x_{\mu}^2} \right)$, and is also a cosmological solution with the cosmological constant not zero.}
\addtocounter{footnote}{1}\footnotetext{Otherwise known as the Schwarzschild-de Sitter solution.}

\begin{center}
{\noindent
\begin{tabular}{|r|l|} \hline \hline
{\bf name [ref.]} & {\bf metric} \\ \hline
N-P-V IIa \cite{kn:NPV,kn:P-VV}  & $-\left(\frac{1-\alpha}{1+\alpha}\right) ^2
\frac{1}{4d^2}\left(\frac{r}{a}\right) ^{2-2n+k} \left[ (c+d)-(c-d)\left(\frac{r}{a}\right) ^{2n} \right] ^{2} dt^2 +$ \\
$0\geq k > -2+\sqrt{2}$  & $ (1+\alpha) ^4\left( \frac{r}{a} \right) ^k (dr^2 + r^2d\Omega ^2)$, \\ 
			 & where $c=2+3k+\frac{3}{4}k^2, d=n(k+2), $ \\
			 & $n = \sqrt{1+2k+\half k^2}$ and $\alpha = \frac{-k}{k+4} = \frac{m}{2a}$\\ \hline
N-P-V IIb \cite{kn:NPV}  & $-\left(\frac{1-\alpha}{1+\alpha}\right) ^2\left(\frac{r}{a}\right) ^{\sqrt{2}}\left[ 1+\frac{\ln a-\ln r}{2\sqrt{2}} \right] ^2 dt^2 +$ \\
$k = -2 + \sqrt{2} $     & $(1+\alpha) ^4\left( \frac{r}{a} \right) ^{-2+\sqrt{2}} (dr^2 + r^2d\Omega ^2)$, \\
			 & where $\alpha = \frac{2-\sqrt{2}}{2+\sqrt{2}} $ \\ \hline
N-P-V IIc \cite{kn:NPV}  & $-\left(\frac{1-\alpha}{1+\alpha}\right) ^2\left(\frac{r}{a}\right) ^{2+k}\cos ^{-2}\Phi \cos ^2\left[ n_1\ln \left(\frac{r}{a}\right) + \Phi \right] dt^2 +$ \\
$-2+\sqrt{2} >k \geq -2$ & $ (1+\alpha) ^4\left( \frac{r}{a} \right) ^k (dr^2 + r^2d\Omega ^2)$, \\ 
			 & where $n_1 = \sqrt{-1-2k-\frac{k^2}{2}}, \tan \Phi = \frac{2+3k+\frac{3k^2}{4}}{n_1(k+2)}$ \\
			 & and $\alpha = \frac{-k}{k+4} = \frac{m}{2a}$ \\ \hline
P-V~\footnotemark IIa \cite{kn:P-V} & $-\left\{ A\cos \left[ \half\mbox{arcsinh}\left(\frac{b^2r^2-c}{\sqrt{b^2-c^2}}\right)+d\right] + B\sin \left[ \half\mbox{arcsinh}\left(\frac{b^2r^2-c}{\sqrt{b^2-c^2}}\right)+d\right] \right\} ^2 dt^2 +$ \\
$b > c$               & $\left( b^2r^4-2cr^2+1 \right) ^{-1} dr^2 + r^2d\Omega ^2$, \\ \hline
P-V IIb \cite{kn:P-V} & $-\left\{ A\cos \left[ \half\mbox{arccosh}\left(\frac{b^2r^2-c}{\sqrt{c^2-b^2}}\right)+d\right] + B\sin \left[ \half\mbox{arccosh}\left(\frac{b^2r^2-c}{\sqrt{c^2-b^2}}\right)+d\right] \right\} ^2 dt^2 +$ \\
$b < c$               & $\left( b^2r^4-2cr^2+1 \right) ^{-1} dr^2 + r^2d\Omega ^2$, \\ \hline
P-V IIc \cite{kn:P-V} & $-\left\{ A\exp \left[ \half\arcsin\left(\frac{b^2r^2-c}{\sqrt{b^2+c^2}}\right)+d\right] + B\exp \left[ -\half\arcsin\left(\frac{b^2r^2-c}{\sqrt{b^2+c^2}}\right)-d\right] \right\} ^2 dt^2 +$ \\
		      & $\left( -b^2r^4-2cr^2+1 \right) ^{-1} dr^2 + r^2d\Omega ^2$, \\ \hline

P-V IV \cite{kn:P-V} & $\frac{r^2}{k^2}\left( A\cos\frac{1}{kr} + B \sin \frac{1}{kr} \right) ^2 dt^2 + \left( k^2r^2 \right) ^{-1} dr^2 + r^2d\Omega ^2$, \\ \hline
P-V V \cite{kn:P-V} & $\left( A\xi ^{\frac{3}{4}}(\xi -1)^{\frac{1}{4}} + B\xi ^{\frac{1}{4}}(\xi -1)^{\frac{3}{4}} \right) ^2 dt^2 + \frac{7}{4(\sqrt{7}r^2+1)} dr^2 + r^2d\Omega ^2$, \\ 
		    & where $\xi = \half (1+\sqrt{1+\sqrt{7}r^2})$ \\ \hline
Wyman~\footnotemark I\cite{kn:Wyman} & $-\left[ AF(a,b;\half;x^2) + BxF(a+\half ,b+\half ;\frac{3}{2};x^2) \right] ^2 dt^2 +$ \\
			& $ x^{-2}dr^2 + r^2d\Omega ^2$, \\ 
			& where $x=\sqrt{1-\left(\frac{r}{R}\right) ^N}, a = \frac{(N-4) + \sqrt{N^2 - 16N + 32} }{4N}$, \\
			& $b = \frac{(N-4) - \sqrt{N^2 - 16N + 32} }{4N}$ and $N > 0$ \\ \hline
Wyman IIa~\footnotemark \cite{kn:Wyman}   & $-\left( Ar^{1-n} - Br^{1+n} \right) ^2 dt^2 +$ \\
$n\neq 2$                   & $\left( \frac{1}{2-n^2} + ar^b[A(2-n) - B(2+n)r^{2n}]^c \right) ^{-1}dr^2 + r^2d\Omega ^2$, \\
$\mbox{and~}n\neq \sqrt{2}$ & $\mbox{where~}b = 2\frac{n^2 - 2}{n-2}, c = 2\frac{2-n^2}{n^2-4}$ \\ \hline
\end{tabular}
}
\end{center}
\addtocounter{footnote}{-2}\footnotetext{P-V IIa,IIb,IIc can be considered as Tolman VII with complex parameters.}
\addtocounter{footnote}{1}\footnotetext{F(a,b;c;d) is the hypergeometric function.}
\addtocounter{footnote}{1}\footnotetext{Wyman II (a,b,c) is a generalization of Tolman VI solution.  As a specific example, Wyman evaluates for $n=1$ while Kuchowicz \cite{kn:Kuchc} also gives specific solutions for $n=\{ \frac{1}{4}, \half ,\frac{3}{4}, \frac{3}{2}\}$.}

\begin{center}
{\noindent
\begin{tabular}{|r|l|} \hline \hline
{\bf name [ref.]} & {\bf metric} \\ \hline
Wyman IIb \cite{kn:Wyman}   & $-\left( \frac{A}{r} - Br^3 \right) ^2 dt^2 + \left( -\half + ar e^{\frac{-A}{4Br^4}} \right) ^{-1}dr^2 + r^2d\Omega ^2$ \\
$n=2$                       & \\ \hline
Wyman ~\footnotemark IIc \cite{kn:Wyman} & $-\left( Ar^{1-\sqrt{2}} - Br^{1+\sqrt{2}} \right) ^2 dt^2 +$ \\
$n=\sqrt{2}$              & $\left( a + \frac{\ln [A(2-\sqrt{2}) - B(2+\sqrt{2})r^{2\sqrt{2}}]}{r(2+\sqrt{2})}\right) ^{-1} dr^2 + r^2d\Omega ^2$ \\ \hline
Wyman~\footnotemark III \cite{kn:Wyman} & $-\left[ cr^{2-n}F\left( \half ,q+1;q+2;(1+2n-n^2)\left( \frac{r}{R}\right) ^N\right) + br^n  \right] ^2 dt^2 +$ \\
			  & $\left[ \frac{1}{1+2n-n^2} - \left(\frac{r}{R} \right) ^N \right] ^{-1}dr^2 + r^2d\Omega ^2$, \\
			  & where $q=\frac{2n}{n^2-2n-1}$ and $N = \frac{2(1+2n-n^2)}{n+1}$ \\ \hline
Wyman IVa \cite{kn:Wyman} & $-\left( \frac{-2r^{2b}\ln r}{c^2(3b-1)} + A r^{2b} + B r^{1-b} \right) dt^2 + $ \\ 
			  & $\left( \frac{-2\ln r}{3b-1} + A c^2 + B c^2r^{1-3b} \right) ^{-1} dr^2 + r^2d\Omega ^2, $ \\
			  & where $b=1+\sqrt{2}$ \\ \hline
Wyman IVb \cite{kn:Wyman} & $-f dt^2 + \left[ a\left(1- \frac{r^2}{R^2} \right) f\right] ^{-1} dr^2 + r^2d\Omega ^2$ \\
			  & where $f = A\left( 1-\sqrt{\frac{R^2}{r^2} -1}\arcsin \frac{r}{R} \right) + \frac{B}{r}\sqrt{1-\frac{r^2}{R^2}} + \frac{1}{a}$, \\
			  & where $a > 0$ and $b > 0$ \\ \hline
Nariai III \cite{kn:Nariai} & $-\left[ \frac{Ar^2+B}{q} \right] ^q dt^2 + \left[ \frac{Ar^2+B}{q} \right] ^{qk} (dr^2 + r^2d\Omega ^2)$, \\ 
			    & where $k\neq -1,k\neq 0 \mbox{~and~} q = \frac{2(1+k)}{k^2-2k-1}$ \\ \hline
Nariai IV \cite{kn:Nariai} & $-A \cos ^2 \left( a-\frac{\sqrt{2}M}{4}r^2\right) \cos ^{-2} \left( b + \frac{M}{4}r^2\right) dt^2 +$ \\
			   & $A \cos ^{-2} \left( b+\frac{M}{4}r^2\right) (dr^2 + r^2d\Omega ^2)$, \\ \hline

Nariai~\footnotemark VI \cite{kn:Nariai} & $-D\left( \frac{ar^p + br^{-p}}{Ar^q+Br^{-q}}\right) ^2 dt^2 + $\\
$\alpha > -1$              & $ D(Ar^{1+q}+Br^{1-q})^{-2} (dr^2 + r^2d\Omega ^2)$, \\ 
			   & where $p = \sqrt{1+\alpha}$ and $q = \sqrt{1+\frac{\alpha}{2}}$ \\ \hline
Nariai VII \cite{kn:Nariai} & $-D\left( \frac{a\cos (p\ln r)+ b\sin (p\ln r)}{Ar^q+Br^{-q}} \right) ^2 dt^2 + $\\
$-1 > \alpha >  -2$         & $D(Ar^{1+q}+Br^{1-q})^{-2} (dr^2 + r^2d\Omega ^2)$, \\ 
			    & where $p = \sqrt{-1-\alpha}$ and $q = \sqrt{1+\frac{\alpha}{2}}$ \\ \hline
Nariai VIII \cite{kn:Nariai} & $-D\left( \frac{a\cos (p\ln r)+ b\sin (p\ln r)}{A\cos(q\ln r)+B\sin (q\ln r)} \right) ^2 dt^2 + $\\
$\alpha < -2 $               & $Dr^{-2}(A\cos(q\ln r)+B\sin (q\ln r))^{-2} (dr^2 + r^2d\Omega ^2)$, \\ 
			     & where $p = \sqrt{-1-\alpha}$ and $q = \sqrt{-\left( 1+\frac{\alpha}{2}\right)}$ \\ \hline
\end{tabular}
}
\end{center}
\addtocounter{footnote}{-2}\footnotetext{Kuchowicz \cite{kn:Kuchc} also corrects Wyman IIc however, neither metric satisfies Einstein's equations; the metric given by Wyman is stated above.}
\addtocounter{footnote}{1}\footnotetext{This is a generalization of Tolman V. Wyman gives the specific example of $n=1$, while Kuchowicz \cite{kn:Kuchc} also gives specific solutions for $n=\{ 2, -\half , 3 \}$.}
\addtocounter{footnote}{1}\footnotetext{Nariai VI through IX resemble those for slowly rotating cylinders.}
\begin{center}
{\noindent
\begin{tabular}{|r|l|} \hline \hline
{\bf name [ref.]} & {\bf metric} \\ \hline
Nariai IX \cite{kn:Nariai} & $-D\left( \frac{a + b\ln r}{Ar^{\frac{1}{\sqrt{2}}}+Br^{-\frac{1}{\sqrt{2}}}} \right) ^2 dt^2 + $\\
$\alpha = -1 $             & $Dr^{-2}\left( Ar^{\frac{1}{\sqrt{2}}} + Br^{-\frac{1}{\sqrt{2}}}\right) ^{-2} (dr^2 + r^2d\Omega ^2)$ \\ \hline
Nariai X \cite{kn:Nariai} & $-D\left( \frac{a\cos \ln r+ b\sin \ln r}{A+B\ln r} \right) ^2 dt^2 + $\\
$\alpha = -2 $            & $Dr^{-2}(A +B\ln r)^{-2} (dr^2 + r^2d\Omega ^2)$ \\ \hline 
Buch1 \cite{kn:Buch1} & $-A\left[ (1+Cr^2)^{3/2} + B\sqrt{2-Cr^2}(5+2Cr^2)\right] ^2 dt^2 + $ \\
		      & $\frac{2(1+Cr^2)}{2-Cr^2} dr^2 +  r^2d\Omega ^2$ \\ \hline
Buch2 \cite{kn:Buch2}  & $-\frac{(1-f)^2}{(1+f)^2}  dt^2 + (1+f) ^4  (dr^2 + r^2d\Omega ^2)$, \\ 
		       & where $f = \frac{a}{2\sqrt{1+kr^2}}$ \\ \hline
Mehra \cite{kn:Mehra}   & $-\left[ \sqrt{1-\frac{16\pi a^2\rho _c}{15}}\cos \left( \frac{z_1 - z}{2}\right) - \frac{2a}{3}\sqrt{\frac{2\pi \rho _c}{5}}\sin \left( \frac{z_1-z}{2}\right) \right] ^{2}  dt^2 +$\\
			& $\left[ 1-\frac{8\pi \rho _c}{15}\left( 5r^2 - \frac{3r^4}{a^2}\right) \right] ^{-1} dr^2 + r^2d\Omega ^2$, \\
			& where $a^2 < \frac{9}{10 \pi \rho _c}, z = \ln \left( \frac{r^2}{a^2} - \frac{5}{6} + \sqrt{\frac{r^4}{a^4} - \frac{5r^2}{3a^2}+\frac{5}{8\pi a^2\rho _c}}\right) $\\
			& and $z_1 = \ln \left( \frac{1}{6} + \sqrt{\frac{5}{8\pi a^2\rho _c} - \frac{2}{3}} \right) $\\ \hline 
Buch3 \cite{kn:Buch3} & $-\frac{a(b-\eta)}{b + \eta} dt^2 + \frac{c(b+\eta )}{a(b - \eta )} dr^2 + \left( \frac{b+\eta}{2a}\right) ^2 r^2 d\Omega ^2$, \\ 
		      & where $\eta = \beta \frac{\sin (Ar)}{Ar}, \beta = \sqrt{b^2-4ac}, a>0, b>0\mbox{~and~}c>0$ \\ \hline
Kucha \cite{kn:Kuchr2,kn:Kucha} & $-\frac{r^2}{R^2}\left[ A\cos \left( m\ln \frac{r}{R}\right) + B\sin \left( m\ln \frac{r}{R}\right) \right] ^2 dt^2 + \left\{ \frac{1}{2+m^2} + \right. $ \\ 
                                & $ \left. C\frac{r^{2a}}{R^{2a}}\left[ (2A+mB)\cos \left( m\ln \frac{r}{R}\right) +(2B-mA)\sin \left( m\ln \frac{r}{R}\right) \right] ^{-a} \right\} ^{-1} dr^2 + $ \\
                                & $r^2d\Omega ^2$, where $a = \frac{2(2+m^2)}{4+m^2}$ \\ \hline
Kuchb Ia \cite{kn:Kuchr2,kn:Kuchb,kn:Kuch} & $-x^{2(1-a)}[A(1+\sqrt{1+bx^2})^{a} + B(1-\sqrt{1+bx^2})^{a}]^2  dt^2 + $\\
$\half < D < 1$                  & $\left( D + Cr^2\right) ^{-1} dr^2 + r^2d\Omega ^2$, \\
				 & where $x \equiv \frac{r}{R}, a = \sqrt{2-\frac{1}{D}}$ and $b = \frac{C}{D}R^2$ \\ \hline
Kuchb Ib \cite{kn:Kuchr2,kn:Kuchb,kn:Kuch} & $-x^2\left[ A + B\ln \left| \frac{1+\sqrt{1+bx^2}}{1-\sqrt{1+bx^2}}\right| \right] ^2  dt^2 + \frac{2}{1+bx^2} dt^2 + r^2d\Omega ^2$, \\ 
$D = \half$                      & where $x = \frac{r}{R}, b = 2CR^2 \neq 0$ \\ \hline 
Kuchb Ic \cite{kn:Kuchr2,kn:Kuchb,kn:Kuch} & $-x^2\left[ A \sin \left( a\ln \left| \frac{1+\sqrt{1+b x^2}}{x}\right| \right) + B \cos \left( a\ln \left| \frac{1+\sqrt{1+b x^2}}{x}\right| \right) \right] ^2  dt^2 + $\\
$0 < D < \half$                  & $\left( D + Cr^2\right) ^{-1} dr^2 + r^2d\Omega ^2$, \\ 
				 & where $x \equiv \frac{r}{R}, a = \sqrt{\frac{1}{D}-2}, b = \frac{C}{D}R^2$ \\ \hline
Kuch1 Ib \cite{kn:Kuch}  & $-\left( Ar + Br\ln r\right) ^2 dt^2 + \frac{1}{C} dr^2 + r^2d\Omega ^2$, \\ \hline
Kuch1 Id \cite{kn:Kuchr2,kn:Kuch}  & $-r^{1+b}\left[ AJ_{\nu}\left( \frac{r^{b-1}}{(b-1)\sqrt{C}}\right) + BJ_{-\nu}\left( \frac{r^{b-1}}{(b-1)\sqrt{C}}\right) \right] ^2 dt^2 +$ \\ 
			 & $ \frac{r^{2(b-1)}}{C} dr^2 + r^2d\Omega ^2$, \\
			 & where $J_{\pm \nu}$ are Bessel functions with $\nu = \frac{\sqrt{(1+b)^2+4b}}{2(b-1)}$ \\ \hline
\end{tabular}
}
\end{center}

\begin{center}
{\noindent
\begin{tabular}{|r|l|} \hline \hline
{\bf name [ref.]} & {\bf metric} \\ \hline
Kuch1 IIIb \cite{kn:Kuch}& $-\left[ Ax^{\sqrt{3}}F\left( \frac{1}{4} + \sqrt{3} + \frac{\sqrt{65}}{4},\frac{1}{4} + \sqrt{3} - \frac{\sqrt{65}}{4};2\sqrt{3};\frac{x}{2-a}\right) + \right. $ \\
			& $ \left. Bx^{1-\sqrt{3}}F\left( \frac{5}{4} - \sqrt{3} + \frac{\sqrt{65}}{4},\frac{5}{4} - \sqrt{3} - \frac{\sqrt{65}}{4};2-2\sqrt{3};\frac{x}{2-a}\right) \right]^2  dt^2 +$ \\
			& $ \left(1  + \frac{2Cr^5}{3} \right) ^{-1} dr^2 + r^2d\Omega ^2$, where $x = \frac{r}{r_b}$ \\ \hline
Kuch1 IV \cite{kn:Kuch}  & $-r^{2(1-\sqrt{2})}e^{C(\sqrt{2} - 1)}\left[ A F\left( \frac{1}{4} - \frac{1}{4\sqrt{2}};\half;\sqrt{2}(2\ln r - C) \right) + \right.  $ \\
			& $\left.  B\sqrt{\ln r -C} F\left( \frac{3}{4} - \frac{1}{4\sqrt{2}};\frac{3}{2};\sqrt{2}(2\ln r-C) \right) \right] ^2  dt^2 + $\\ 
			& $\left( C - 2\ln r \right) ^{-1} dr^2 + r^2d\Omega ^2$ \\ \hline
Kuch1 V \cite{kn:Kuchr2,kn:Kuch}   & $-r^{2d}\left( A F(\alpha ,\beta ;\gamma ;-bCr^{2b}) + \right. $  \\
			& $\left. Br^{2b(1-\gamma)}F(\alpha - \gamma +1,\beta -\gamma +1;2-\gamma ;-bCr^{2b}) \right) ^2 dt^2 + $\\
			& $\left( \frac{1}{b} + Cr^{2b}\right) ^{-1} dr^2 + r^2d\Omega ^2$, \\
			& where $d=b+\sqrt{b^2 - b + 1} , \gamma = 1 + \frac{1}{b}\sqrt{b^2-b+1}, \gamma \nel$ I \\
			& $\alpha = \frac{1}{2b}\left( \frac{3}{2}b-1+\sqrt{b^2-b+1} + \sqrt{\frac{b^2}{4} - 2b + 2} \right)$ and \\
			& $\beta = \frac{1}{2b}\left( \frac{3}{2}b-1+\sqrt{b^2-b+1} - \sqrt{\frac{b^2}{4} - 2b + 2} \right)$ \\ \hline
Kuch2~\footnotemark I \cite{kn:Kuchr2,kn:Kuch2}  & $-Br^A dt^2 + \left( Cr^{-a} - \frac{4}{A^2-4A-4} \right) ^{-1} dr^2 + r^2d\Omega ^2$, \\ 
			 & where $a=\frac{A^2-4A-4}{A+2}$ with $A \neq -2$ and $A \neq 2(1 \pm \sqrt{2})$ \\ \hline
Kuch2 III \cite{kn:Kuchr2,kn:Kuch2}& $-Be^{\frac{Ar^2}{2}} dt^2 + \left\{ 1+r^2e^{-\half Ar^2}\left[ C - \frac{A}{2e} \mbox{Ei} \left( 1+\frac{Ar^2}{2}\right) \right] \right\} ^{-1} dr^2 + r^2d\Omega ^2$ \\ \hline
Kuch2 IV \cite{kn:Kuch2} & $-Be^{\frac{x}{2}} dt^2 + \frac{x |x+2|^4}{Ce^{\frac{x}{2}} + 8(x^4+14x^3+96x^2+384x+784)} dr^2 + r^2d\Omega ^2$, \\ 
			 & where $x = \frac{A}{r^2}$ \\ \hline
Kuch2 VI \cite{kn:Kuchr2,kn:Kuch2} & $-B(r+A)^2 dt^2 + \left[ \left( C - \frac{4}{A^2} \right) r^2 - \frac{2r}{A}+ 1 + \frac{4r^2}{A^2} \ln \left| 1+\frac{A}{2r} \right| \right] ^{-1} dr^2 + $ \\
			 & $r^2d\Omega ^2$ \\ \hline
Kuch2~\footnotemark VII \cite{kn:Kuch2}& $-Ar^{2(1\pm\sqrt{2})} dt^2 + \left( C - \frac{2}{2\pm\sqrt{2}}\ln r \right) ^{-1} dr^2 + r^2d\Omega ^2$ \\ \hline
Whittaker \cite{kn:Whittaker} & $-n\left( 1+B-\frac{B}{a r}\sqrt{1-a ^2r^2}\arcsin (a r)\right) dt^2 + \left[ (1+B)(1-a ^2r^2)- \right. $ \\
			  & $\left. \frac{B}{a r}(1-a ^2r^2)^{3/2}\arcsin (a r)\right] ^{-1} dr^2 + r^2 d\Omega ^2$, \\
			  & where $B = \frac{C}{2a ^2}$ and $ a ^2 = \half nl $ \\ \hline
B-L \cite{kn:B-L}& $-\frac{\rho _br^2}{3} dt^2 + \left( \half - \frac{\rho _br^2}{6}\right) ^{-1} dr^2 + r^2d\Omega ^2$ \\ \hline
Kuch68 I \cite{kn:Kuch68} & $-\left( A\sqrt{1+\frac{a}{r}} + B\left[ \frac{r^2}{2a^2} - \frac{5r}{4a} - \frac{15}{4} + \frac{15}{8}\sqrt{1+\frac{a}{r}}\ln \left( \frac{2r}{a} + 1 + \frac{2}{a}\sqrt{r(r+a)}\right) \right] \right) ^2 dt^2 + $ \\
                          & $  \left( 1+\frac{a}{r} \right) ^{-1} dr^2 + r^2d\Omega ^2$ \\ \hline
\end{tabular}
}
\end{center}
\addtocounter{footnote}{-1}\footnotetext{Kuch2 I is only a slightly more general version of Tolman V, obtained 
by setting $C=1$.  There are many metrics which are interconnected with one another.  In this paper we have noted some
generalizations and duplications but we have not exhausted all the possible interconnections between metrics.}
\addtocounter{footnote}{1}\footnotetext{Kuchowicz does not state that Kuch2 II is a special case of Kuch2 VII.}

\begin{center}
{\noindent
\begin{tabular}{|r|l|} \hline \hline
{\bf name [ref.]} & {\bf metric} \\ \hline
Kuch68 II \cite{kn:Kuch68} & $-\left( 1+\frac{a}{r} \right) dt^2 + \left[ \left( 1+\frac{a}{r} \right)(1+C(a+2r)^2) \right] ^{-1} dr^2 + r^2d\Omega ^2$ \\ \hline
Leib I \cite{kn:Leib}  & $-(A-Bx)^2 dt^2 + \frac{F(x)}{x^2F(x) + ar^2} dr^2 + r^2d\Omega ^2$, \\
		       & where $F(x) = [x(A-Bx) + B(1-x^2)]^2\left[\frac{(x-y)^{y}}{(x-z)^{z}}\right] ^{2/(z-y)}$, \\ 
                       & $x = \sqrt{1-\frac{r^2}{R^2}}, y = \frac{A-\sqrt{A^2+8B^2}}{4B}$ and $ z = \frac{A+\sqrt{A^2+8B^2}}{4B}$ \\ \hline
Leib IV \cite{kn:Leib} & $-g^{-1} dt^2 + \frac{(2-ar)^6}{r^2g^3\left[ 4a^3r+\frac{64}{r^2}+32a^2\ln \left(\frac{g}{r^2}\right)-\frac{4a^2}{g} + b\right]} dr^2 + r^2d\Omega ^2$, \\ 
		       & where $g = 1 - ar $ \\ \hline
Heint IIa \cite{kn:Heint}  & $-A^2\left( 1+ ar^2 \right) ^3 dt^2 +  \left( 1-\frac{3ar^2}{2}\frac{1+C(1+4ar^2) ^{-\half}}{1+ar^2} \right) ^{-1} dr^2 + r^2d\Omega ^2$ \\ \hline
Heint IIb \cite{kn:Heint}  & $-B^2\left( 5+ 2ar^2 \right) ^2\left( 2-ar^2\right) dt^2 + $ \\
			   & $ \left( 1-\frac{3ar^2}{2}\frac{1+C(5-4ar^2)}{1+ar^2} \right) ^{-1} dr^2 + r^2d\Omega ^2$ \\ \hline
Heint IIIa \cite{kn:Heint} & $-( a+br^{-2})^2 dt^2 + $ \\ 
			   & $ \left[ 1-2r^2\frac{2b^2}{\left( b-ar^2\right) ^2}\left(\frac{6}{r^2} - \frac{a}{b} + Cr^4\right) \right] ^{-1} dr^2 + r^2d\Omega ^2$ \\ \hline
Heint IIIb \cite{kn:Heint} & $-\left( a+\frac{b}{r} \right) ^2 dt^2 + $ \\
			   & $\left[ 1-\frac{3r^2}{2}\left( \frac{1}{r^2} - \frac{a}{2br} - \frac{a^2}{8b^2} + C e^{\frac{-4b}{ar}} \right) \right] ^{-1} dr^2 + r^2d\Omega ^2$ \\ \hline
Heint IIIe \cite{kn:Heint} & $-\left( a+b r^3 \right) ^2 dt^2 +  \left[ 1-2r^2\left( \frac{3br + C}{4br^3 + a}\right) \right] ^{-1} dr^2 + r^2d\Omega ^2$ \\ \hline
Kuch3 Ia \cite{kn:Kuch3}  & $-\left( Ar^{\alpha} + Br^{\beta}\right) ^2 dt^2 + \left\{-\frac{1}{D(1+\alpha )} + \right. $ \\
$B > 0$                   & $ \left. r^{-2D}\left( C-\frac{4}{(1+\alpha )^{3/2}(1+\beta )^{1/2}}\sqrt{\frac{B}{A}}\arctan \left[\sqrt{\frac{(1+\alpha )A}{(1+\beta )B}}r^{2D} \right]\right)\right\} ^{-1} dr^2 + r^2d\Omega ^2 $, \\ 
			  & where $D = \frac{1}{15}(4+2\sqrt{34}), \alpha = \frac{1}{3}(5+\sqrt{34})$ and $ \beta = \frac{1}{5}(3-\sqrt{34}) $ \\ \hline
Kuch3 Ib \cite{kn:Kuch3}  & $-\left( Ar^{\alpha}\right) ^2 dt^2 + \left( Cr^{-2D} - \frac{1}{D(1+\alpha )} \right) ^{-1} dr^2 + r^2d\Omega ^2$, \\ 
$B = 0$                   & where $D = \frac{1}{15}(4+2\sqrt{34})$ and $ \alpha = \frac{1}{3}(5+\sqrt{34}) $ \\ \hline
Kuch3 Ic \cite{kn:Kuch3}  & $-\left( Ar^{\alpha} + Br^{\beta}\right) ^2 dt^2 + $ \\
$B < 0$                   & $ \left\{-\frac{1}{D(1+\alpha )} + r^{-2D}\left( C-\frac{4}{(1+\alpha )^{3/2}(1+\beta )^{1/2}}\sqrt{\frac{-B}{A}}\ln \Phi (r)\right)\right\} ^{-1} dr^2 + r^2d\Omega ^2$, \\ 
			  & where $\Phi (r) = \frac{(1+\beta )B - (1+\alpha )Ar^{4D}+2\sqrt{-AB(1+\alpha )(1+\beta )}r^{2D}}{(1+\beta )B+(1+\alpha )Ar^{4D}}, $ \\ 
			  & $D = \frac{1}{15}(4+2\sqrt{34}), \alpha = \frac{1}{3}(5+\sqrt{34})$ and $ \beta = \frac{1}{5}(3-\sqrt{34}) $ \\ \hline
Kuch3 II \cite{kn:Kuch3} & $-\left( Ar^{\alpha} + Br^{\beta}\right) ^2 dt^2 + \left\{ - \frac{1}{D(1+\alpha )}+r^{-2D}\left( C-\frac{2B^{1/3}}{A^{1/3}(1+\alpha )^{4/3}(1+\beta )^{2/3}} \right. \right. \times $ \\
			 & $\left.\left.\left[ \frac{3}{2}\ln \frac{\sqrt[3]{A(1+\alpha )}r^{2D} + \sqrt[3]{B(1+\beta )}}{\sqrt[3]{A(1+\alpha ) r^{6D}+B(1+\beta )}} 
			   -\sqrt{3}\arctan\frac{\sqrt{3}r^{2D}}{r^{2D}-2\sqrt[3]{\frac{B(1+\beta )}{A(1+\alpha )}}} \right]\right)\right\} ^{-1} dr^2 +$\\
			 & $ r^2d\Omega ^2$, where $D = \frac{4+2\sqrt{74}}{35}, \alpha = \frac{1}{5}(7+\sqrt{74})$ and $ \beta = \frac{1}{7}(5-\sqrt{74}) $ \\ \hline
\end{tabular}
}
\end{center}

\begin{center}
{\noindent
\begin{tabular}{|r|l|} \hline \hline
{\bf name [ref.]} & {\bf metric} \\ \hline
Kuch3 III \cite{kn:Kuch3} & $-\left( Ar^{\alpha} + Br^{\beta}\right) ^2 dt^2 + \left\{ Cr^{-2D} - \frac{2n}{D[4n+(n-2)D]} + \frac{8n^2r^{-2D}}{4n+(n-2)D} \times \right. $ \\
			  & $\left.\left[ {\displaystyle \sum_{k=1}^{n} }\frac{(-1)^{k-1}}{n+1-k} \left[\frac{A}{B}\right] ^{k-1}\frac{[4n+(n-2)D]^{k-1}}{[4n+(n+2)D]^k}r^{\frac{2D}{n}(n+1-k)} + \right.\right. $ \\
			  & $\left.\left. (-1)^n\left[ \frac{A}{B} \right] ^n\frac{[4n+(n-2)D]^n}{[4n+(n+2)D]^{n+1}} \ln \left( \frac{A[4n+(n-2)D]}{2n} + \frac{B[4n+(n+2)D]r^{\frac{2D}{n}}}{2n}\right)\right]\right\} ^{-1} dr^2+ $\\
			  & $ r^2d\Omega ^2$, where $\alpha = 1+D(\frac{1}{2} -n), \beta = 1+D(\frac{1}{2} +n)$, \\ 
			  & $D_{\pm} = \frac{-4n^2\pm 2n\sqrt{2(n^2+4)}}{n^2-4}$ and $n \in I > 3$ \\ \hline
Kuch3 IV \cite{kn:Kuch3} & $-r^{2+D}( A + B\ln r) ^2 dt^2 + \left\{ Cr^{-2D} - \frac{1}{D(1+\alpha )} + \frac{2}{(1+\alpha )^2} \times \right. $ \\
			 & $\left.\exp \left[ -2D \left(\frac{A}{B} + \frac{1}{1+\alpha}\right) \right] r^{-2D}\mbox{Ei} \left[ 2D\left( \frac{A}{B} + \frac{1}{1+\alpha}\right) + 2D\ln r \right] \right\} ^{-1} dr^2 + $\\
			 & $+ r^2d\Omega ^2$, where $D_{\pm} = \pm 2\sqrt{2}-4$ and $\alpha = 1+D/2$ \\ \hline
Kuch3 V \cite{kn:Kuch3} & $-B^2r^2e^{2Ar} dt^2 + [F(r)]^{-1} dr^2 + r^2d\Omega ^2 $ \\
			& where $F(r) = Cr^2e^{-Ar} + \frac{1+Ar}{2} + A^2r^2e^{-2Ar}\left( \frac{1}{2e^4}\mbox{Ei}(2Ar+4) - \frac{5}{4}\mbox{Ei} (2Ar)\right) $\\ \hline
Kuch71 I  \cite{kn:Kuch71}& $-\left[ (ar^4)^{\frac{(1-b)}{2-k}}\left( \xi ^{1-2m}\frac{d}{d\xi}\right)^{n+1}\left(Ae^{\frac{\sqrt{c}}{m}\xi ^m} + Be^{-\frac{\sqrt{c}}{m}\xi ^m} \right) \right] ^2 dt^2 + $ \\ 
			  & $ \left[ \frac{(k-2)^2}{16}(ar^{2k})^{\frac{2}{k-2}}\right] ^{-1} dr^2 + r^2d\Omega ^2$, where $\xi = (ar^4)^{\frac{1-2b}{2-k}}$, \\
			  & $b=\half \pm \frac{2n+1}{2\sqrt{2n^2+2n+1}}, k = \mp \frac{2}{\sqrt{2n^2+2n+1}}, c = \frac{a}{2b-1}$ and $ m = \frac{1}{2n+1} $ \\ \hline
Kuch71 II \cite{kn:Kuch71}& $-\left( Ae^{\frac{1+\sqrt{2}}{4}Ar^2} + Be^{\frac{1-\sqrt{2}}{4}Ar^2}\right) ^2 dt^2 + Ce^{\half Ar^2} (dr^2 + r^2d\Omega ^2)$ \\ \hline
Kuch71b \cite{kn:Kuch71b} & $-ar^{2\sqrt{2s^2-\half}} dt^2 + \frac{br^{4s-2-2\sqrt{2s^2-\half}}}{cr^{8s} + 2\sqrt{cd}r^{4s}+d} (dr^2 + r^2d\Omega ^2)$ \\ \hline
Kuch5 I \cite{kn:Kuch5}  & $-r^4e^{Ar(1-\sqrt{2})}\left\{ C F(\delta ;3;A\sqrt{2}r)+D\left[ -\frac{1}{2A^2r^2}+\frac{2-\sqrt{2}}{4Ar}+ \right. \right.  $ \\
			 & $\left. \left. \half (\delta -1)(\delta -2) F(\delta ;3;A\sqrt{2}r) \times\ln (A\sqrt{2}r) + \right.\right.$ \\
			 & $\left. \left. + {\displaystyle \sum^{\infty}_{k=1}} \frac{(\delta -2)(\delta -1)\delta (\delta +1)...(\delta +k-1)}{2\cdot 3\cdot 4...(2+k)\cdot k!}(A\sqrt{2}r)^k\times 
			   {\displaystyle \sum^{k-1}_{v=0}} \left(\frac{1}{\delta +v}-\frac{1}{3+v}-\frac{1}{1+v}\right)\right]\right\} ^2 dt^2 + $ \\
			 & $Be^{Ar} (dr^2 + r^2d\Omega ^2)$, where $\delta = \half (3+\sqrt{2})$ \\ \hline
Kuch5 IV \cite{kn:Kuch5} & $-r^4e^{-\sqrt{2}Ar^4}\left[ C F\left(\frac{3-\sqrt{2}}{4};\frac{3}{2};\frac{A}{\sqrt{2}}r^4\right) +\frac{D}{r^2} F\left(\frac{1-\sqrt{2}}{4};\half ;\frac{A}{\sqrt{2}}r^4\right) \right] ^2 dt^2 + $ \\
			 & $ Be^{Ar^2} (dr^2 + r^2d\Omega ^2)$ \\ \hline
Kuch5 VIII \cite{kn:Kuch5} & $-Ce^{ar^2} dt^2 + De^{-ar^2}\coth ^{-2}\left( -\frac{a\sqrt{2}}{2}r^2+A\right) (dr^2 + r^2d\Omega ^2)$ \\ \hline
Kuch5 IX \cite{kn:Kuch5}   & $-Ar^{2\sqrt{2s^2-\half}} dt^2 + \frac{r^{4s-2-2\sqrt{2s^2-\half}}}{(Br^{4s}-C)^2} (dr^2 + r^2d\Omega ^2)$, where $s > \frac{1}{2}$ \\ \hline
Kuch5 XI \cite{kn:Kuch5}   & $-CF(r) dt^2 + D\left(B+\frac{r^{2-2A}}{2(A-1)}\right) ^{-2}F(r) (dr^2 + r^2d\Omega ^2)$, \\ 
			   & where $F(r) = \exp \left( 2\sqrt{\frac{2A}{1-A}}\arctan \sqrt{2B(1-A)r^{2A-2}-1}\right)$ and $A \neq 1$ \\ \hline 
Kuch5 XII \cite{kn:Kuch5}  & $-C\left(\frac{\sqrt{2A}F(r)-1}{\sqrt{2A}F(r)+1}\right) ^{\sqrt{2}} dt^2 + \frac{D e^{2Ar^2}}{[F(r)]^2}\left( \frac{\sqrt{2A}F(r)+1}{\sqrt{2A}F(r)-1}\right) ^{\sqrt{2}} (dr^2 + r^2d\Omega ^2)$, \\ 
			   & where $F(r) = \frac{1}{2A} + Be^{Ar^2}$ \\ \hline
Kuch5 XIII \cite{kn:Kuch5} & $-C\left| (a^2-\half )r^2+B\right| ^{\frac{2a}{2a^2-1}} dt^2 + D\left| (a^2-\half )r^2+B\right| ^{\frac{2-2a}{2a^2-1}} (dr^2 + r^2d\Omega ^2)$ \\ \hline
\end{tabular}
}
\end{center}

\begin{center}
{\noindent
\begin{tabular}{|r|l|} \hline \hline
{\bf name [ref.]} & {\bf metric} \\ \hline
Kuch5 XV \cite{kn:Kuch5}   & $-C\left| \sqrt{2}F(r)+Be^{Ar^2}-A \right| ^{-\frac{1}{\sqrt{2}}} e^{\frac{1}{A}F(r)} dt^2 + $ \\ 
			   & $ D\left| \sqrt{2}F(r)+Be^{Ar^2}-A \right| ^{\frac{1}{\sqrt{2}}} e^{\frac{1}{A}[Be^{Ar^2}-F(r)]} (dr^2 + r^2d\Omega ^2)$, \\ 
			   & where $F(r) = \sqrt{\half B^2e^{2Ar^2}-ABe^{Ar^2}}$ \\ \hline
Kuch5 XVI \cite{kn:Kuch5}  & $-C\left| B\sqrt{2}F(r)+B^2r^2 + AB\right| ^{-\frac{1}{\sqrt{2}}} e^{\frac{A+Br^2}{2B}F(r)} dt^2 + $ \\
			   & $ D\left| B\sqrt{2}F(r)+B^2r^2 + AB\right| ^{\frac{1}{\sqrt{2}}} e^{Ar^2+\half Br^4 - \frac{A+Br^2}{2B}F(r)} (dr^2 + r^2d\Omega ^2)$, \\ 
			   & where $F(r) = \sqrt{\frac{B^2}{2}r^4 + ABr^2 + \frac{A^2}{2} - B}$ \\ \hline
R-R I \cite{kn:R-R}    & $-\frac{1}{C^2} dt^2 + \frac{C^4}{\left( 1 - \frac{r^2}{R^2} \right)} dr^2 + C^4r^2d\Omega ^2$ \\ \hline
R-R II \cite{kn:R-R}   & $-f^{-1} dt^2 + C^2f dr^2 + f^2 r^2d\Omega ^2$, \\ 
		       & where $f = C^2\left(1-\frac{2m}{r} - \frac{r^2}{R^2}\right)$ \\ \hline
R-R III \cite{kn:R-R}  & $-f^{-1} dt^2 + f^2 \left( 1 - \frac{r^2}{R^2} \right) ^{-1}  dr^2 + f^2r^2d\Omega ^2$, \\ 
		       & where $f = \left(A-B\sqrt{1- \frac{r^2}{R^2}}\right) ^2$ \\ \hline
R-R IV \cite{kn:R-R}  & $-f^{-1} dt^2 + B^2f  \frac{1 + \frac{2r^2}{A^2}}{1-\frac{r^2}{R^2}} dr^2 + f^2r^2d\Omega ^2$, \\ 
		      & where $f = B^2\left( 1 + \frac{r^2}{A^2}\right)$ \\ \hline
R-R V \cite{kn:R-R}   & $-\frac{1}{B^2r} dt^2 + B^4r^2\frac{7}{4-7\left( \frac{r}{R} \right) ^{7/3}} dr^2 + B^4r^4d\Omega ^2$ \\ \hline
R-R VI \cite{kn:R-R}  & $-f^{-1} dt^2 + f^2(2-n^2)  dr^2 + f^2r^2d\Omega ^2$, \\ 
		      & where $f=(Ar^{1-n} - B^{1+n}) ^2$ \\ \hline
R-R VII \cite{kn:R-R} & $-f^{-1} dt^2 + f^2 \left( 1 - \frac{r^2}{R^2} + \frac{4r^4}{A^4} \right) ^{-1} dr^2 + f^2 r^2d\Omega ^2$, \\ 
		      & where $f=B^2\sin ^2 \ln \sqrt{ \frac{\sqrt{1-\frac{r^2}{R^2}+\frac{4r^4}{A^4}} + \frac{2r^2}{A^2} - \frac{A^2}{4R^2}}{C}}$ \\ \hline
R-R VIII \cite{kn:R-R} & $-f^{-1} dt^2 + B^2r^{2b}f dr^2 + f^2 r^2d\Omega ^2$, \\ 
		       & where $f = B^2r^{2b}\left( \frac{2}{(a-b)(a+2b-1)} - \left( \frac{2m}{r}\right) ^{(a + 2b -1)} - \left( \frac{r}{R} \right) ^{(a-b)}\right) $\\ 
		       & and $(a+b)(a-1) - 2b -2 = 0 $ \\ \hline
Kuch73 I \cite{kn:Kuch73} & $-\frac{ar^{2}}{(1+cr^{\sqrt{2}})^2} dt^2 + \frac{br^{\sqrt{2}-2}}{(1+cr^{\sqrt{2}})^2} (dr^2 + r^2d\Omega ^2)$ \\ \hline
Kuch73 II \cite{kn:Kuch73} & $-\frac{a}{c-r^2} dt^2 + \frac{b}{|c-r^2|^{\sqrt{3}}(1-y)^2} (dr^2 + r^2d\Omega ^2)$ \\
                           & where $y=\frac{d}{(c-r^2)^{\sqrt{3}}}$  \\ \hline
K-N-B \cite{kn:K-N-B} & $-c\exp\left[ \sin ^{-1} \left( \frac{a^2r^2 +2b}{2\sqrt{a^2+b^2}} \right) \right] dt^2 + \left( 1-br^2-\frac{a^2}{4}r^4 \right)^{-1} dr^2 + r^2d\Omega ^2$ \\ \hline
%
%
G-G \cite{kn:G-G} & $-A\left( \frac{2r^2-(b^2d+1+c)}{2r^2 - (b^2d+ 1 - c)}\right) ^{\frac{2b}{c}} dt^2 + $ \\
		  & $ \frac{B}{b^2(2+d)-(b^2d+ 1)r^2 + r^4}\left( \frac{2r^2-(b^2d+1+c)}{2r^2 - (b^2d+ 1 - c)}\right) ^{-(b^2d+ 2b-1)/c} (dr^2+r^2d\Omega ^2)$, \\ 
		  & where $c= \sqrt{(b^2d-1)^2-8b^2} $ \\ \hline
\end{tabular}
}
\end{center}

\begin{center}
{\noindent
\begin{tabular}{|r|l|} \hline \hline
{\bf name [ref.]} & {\bf metric} \\ \hline
Bayin~\footnotemark III \cite{kn:Bayin} & $-A^2 \left( r + \sqrt{r^2 - \frac{2}{B}} \right) ^{2/\sqrt{B}}  dt^2 + \left( \frac{2}{r^2} - B \right) ^{-1}  dr^2 + r^2d\Omega ^2$ \\ \hline
Bayin VI \cite{kn:Bayin}  & $-AC^{-a}e^{-dar^2}  dt^2 + BC^{b}e^{dbr^2} (dr^2 + r^2d\Omega ^2)$, \\
			  & where $a^2+2ab-b^2=0$ \\ \hline
Gold I \cite{kn:Gold} & $-Af^{\frac{1}{b}} dt^2 + B\frac{g^2}{c^2(1+2ag + 2g^2)}f^{-\frac{1+a}{b}} (dr^2 + r^2d\Omega ^2)$, \\
		      & where $f = \frac{1+(a+b)g}{1+(a-b)g}, g =\frac{c}{1-dr^2}$ and $b= \sqrt{a^2 - 2}$ \\ \hline
Gold II \cite{kn:Gold} & $-A\left[ \frac{(1+c)u^2+1-c}{2u}\right] ^{\sqrt{2}c/(1-c^2)}u^{\frac{-\sqrt{2}}{1-c^2}} dt^2 + $ \\
		       & $ B\left[ \frac{2u}{(1+c)u^2+1-c}\right] ^{(2+\sqrt{2}c)/(1-c^2)}u^{\frac{\sqrt{2}+2c}{1-c^2}} (dr^2 + r^2d\Omega ^2)$,\\
		       & where $u =e^{a-b r^2}$ \\ \hline
Gold III \cite{kn:Gold} & $-A\left( \frac{g-1}{g+1}\right) dt^2 + B\left( 1+\frac{1}{g}\right) ^2 (dr^2 + r^2d\Omega ^2)$,\\
			& where $g = \cosh (a+br^2)$ \\ \hline
M-W I \cite{kn:M-W} & $-\left( 1+a\frac{r^2}{R^2}\right) (A\sin z +B\cos z )^2 dt^2 + \left( 1+2a\frac{r^2}{R^2} \right) dr^2 + r^2d\Omega ^2,$ \\
                    & where $z=\sqrt{1+2a\frac{r^2}{R^2}} - \tan ^{-1} \sqrt{1+2a\frac{r^2}{R^2}} $ \\ \hline
M-W II \cite{kn:M-W} & $-\left( A\left[ \half+a\frac{r^2}{R^2} - \sqrt{1+a\frac{r^2}{R^2}}\cosh ^{-1}\sqrt{\half \left( 1+a\frac{r^2}{R^2}\right)} \right] +B\sqrt{1+a\frac{r^2}{R^2}}  \right) ^2 dt^2 + $ \\
                     & $\frac{1+2a\frac{r^2}{R^2}}{1+a\frac{r^2}{R^2}} dr^2 + r^2d\Omega ^2$ \\ \hline
M-W III \cite{kn:M-W} & $-\left(\frac{2r}{3R} - \frac{r^2}{9R^2}\right) dt^2 + \frac{63}{36-\frac{r^2}{R^2}} dr^2 + r^2d\Omega ^2$ \\ \hline
K-K \cite{kn:K-K} & $-\left( \frac{r}{R} \right)^{2A}\left( 1+ar^{-k}\right)^{1-2A} dt^2 + (1-2A-A^2)r^k\frac{r^k+b}{(r^k+a)^2} dr^2 + r^2d\Omega ^2$ \\ \hline
Stewart \cite{kn:Stew} & $-\left( 1-\frac{M}{2a} \right) ^2\left( 1-\frac{Mr^2}{2a^3} \right) ^2\left( 1+\frac{M}{a} - \frac{Mr^2}{a^3} - \frac{M^2r^2}{4a^4} \right) ^{-2} dt^2 +$ \\ 
		       & $ \left( 1-\frac{M}{2a} \right) ^2\left( 1-\frac{Mr^2}{2a^3} \right) ^{-6}\left( 1+\frac{M}{a} - \frac{Mr^2}{a^3} - \frac{M^2r^2}{4a^4} \right) ^{4}  (dr^2 + r^2d\Omega ^2)$ \\ \hline
P-S Ia \cite{kn:P-S} & $-B^2r^{2(c-b)}(1-kr^{4b}) dt^2 + \frac{1+c-b-(1+c+b)kr^{4b}}{(1-kr^{4b})\left( \frac{A}{r^a} - \frac{2}{a}\right) }dr^2 + r^2d\Omega ^2$, \\
		     & where $b=\frac{\sqrt{a^2+16a+32}}{4}, c=\frac{a+4}{4}$ and $ a \neq 0 $ \\ \hline
P-S Ib \cite{kn:P-S}  & $-B^2r^{2(1-\sqrt{2})}(1-kr^{4\sqrt{2}}) dt^2 + \frac{2-\sqrt{2}-(2+\sqrt{2})kr^{4\sqrt{2}} }{ (1-kr^{4\sqrt{2}})(A - 2\ln r) }dr^2 + r^2d\Omega ^2$ \\ \hline
Durg~\footnotemark IV \cite{kn:Durg} & $-A(1+Cr^2)^4 dt^2 + \left( \frac{7-10Cr^2-C^2r^4}{7(1+Cr^2)^2} + \frac{KCr^2}{(1+Cr^2)^2(1+5Cr^2)^{\frac{2}{5}}}\right) ^{-1} dr^2 + r^2d\Omega ^2$ \\ \hline
Durg V \cite{kn:Durg}  & $-A(1+Cr^2)^5 dt^2 + \left( \frac{1-\frac{Cr^2(309+54Cr^2+8C^2r^4)}{112} + \frac{KCr^2}{\sqrt[3]{1+6Cr^2}}}{(1+Cr^2)^3} \right) ^{-1} dr^2 + r^2d\Omega ^2$ \\ \hline
\end{tabular}
}
\end{center}

\addtocounter{footnote}{-1}\footnotetext{Bayin III may be contained within Kuch1 V and Bayin VI within Kuch5 IV  although the specific values of the hypergeometric equation are not given.}
\addtocounter{footnote}{1}\footnotetext{The general form was originally found by Korkina \cite{kn:Kork} but the only solution he presented was Heint IIa.}

\begin{center}
{\noindent
\begin{tabular}{|r|l|} \hline \hline
{\bf name [ref.]} & {\bf metric} \\ \hline
Whitman II \cite{kn:Whit2}  & $-\left( A^2 \left( 1+a\frac{r^2}{R^2} \right) + B\frac{r}{R\sqrt{2a}}\sinh \beta \left[ 1 + \frac{R^2+2ar^2}{2ar^2} \right. \right.$ \\
                           & $\left.\left.\times \left( 1 - \sqrt{10a}\frac{r}{\sqrt{R^2+2ar^2}}\coth \beta \right) \right] \right) dt^2 + B^2\left(1+2a\frac{r^2}{R^2}\right)$\\
                           & $\left( A^2 \left( 1+a\frac{r^2}{R^2} \right) + B\frac{r}{R\sqrt{2a}} \sinh \beta \right. $\\
                           & $\left. \times \left[ 1 + \frac{R^2+2ar^2}{2ar^2}\left( 1-\sqrt{10a}\frac{r}{\sqrt{R^2+2ar^2}}\coth \beta  \right)\right] \right) ^{-1} dr^2 + r^2d\Omega ^2$ \\
                           & where $ \beta = \sqrt{5}\mbox{arcsinh} \sqrt{\frac{R^2}{2ar^2}} $ \\ \hline
Whitman III \cite{kn:Whit2}  & $- f dt^2 + af\left( 1-b\frac{r^2}{R^2}\right) ^{-9}  dr^2 + r^2d\Omega ^2 $\\ 
                            & where $ f = \frac{r^2}{R^2}\left( a\left( 1-\frac{30br^2}{7R^2} + \frac{32b^2r^4}{7R^4} + \frac{64b^3r^6}{49R^6} - \frac{640b^4r^8}{147R^8} + \frac{256b^5r^{10}}{147R^{10}}\right) +  \right.  $\\
                            & $\left. + A\left( 1-b\frac{r^2}{R^2} \right) ^{-4} \left[1\left( 1-b\frac{r^2}{R^2}\right) \left( 1-2b\frac{r^2}{R^2}\right) + \frac{\sqrt{R^2-br^2}}{\sqrt{b}r}\arctan \frac{\sqrt{b}r}{\sqrt{R^2-br^2}}\right] \right)  $\\ \hline
Whitman IV \cite{kn:Whit2}  & $-f  dt^2 + \frac{a}{f}\left( 1-b\frac{r^2}{R^2} \right)^3 dr^2 + r^2d\Omega ^2 $\\
                             & where $f= \frac{r^2}{R^2} (g_p + g_h)$, $g_p = a\frac{R^2}{r^2} \left( 1-\frac{3}{8}\frac{br^2}{R^2} - \frac{3}{4}\frac{b^2r^4}{R^4} + \frac{1}{2}\frac{b^3r^6}{R^6} \right), $ \\
                             & $g_h = C\left( 1-b\frac{r^2}{R^2}\right) ^{\frac{3}{2}} \left( \frac{3}{4} \frac{\cos \beta}{\left( 1-b\frac{r^2}{R^2}\right) ^{\frac{3}{2}}} - \frac{3\sqrt{3}}{2} \frac{\sin \beta}{\left( 1-b\frac{r^2}{R^2}\right)\sqrt{\frac{br^2}{R^2}}} - 3R^2 \frac{\cos \beta}{br^2\sqrt{1-b\frac{r^2}{R^2}}} + \frac{\sqrt{3}}{2} \frac{\sin \beta}{\left( \frac{br^2}{R^2}\right) ^{\frac{3}{2}}} \right) $ \\
                             & and $ \beta = 2\sqrt{3} \arcsin \sqrt{\frac{br^2}{R^2}}  $\\ \hline
D-P-P\footnotemark ~ I \cite{kn:D-P-P}  & $-\frac{B}{\sqrt{1-cr^2}} dt^2 + \frac{(2-cr^2)^4}{4(1-cr^2)^2(4-4cr^2-c^2r^4) + 16Acr^2(1-cr^2)^{\frac{5}{2}}} dr^2 + r^2d\Omega ^2$ \\ \hline
V-P II \cite{kn:V-P}  & $-F^2 \cos ^{-2}\ln (C\sqrt{1-Ar^2}) dt^2 + (1-Ar^2)^{-2} dr^2 + r^2d\Omega ^2$ \\ \hline
P-S2 \cite{kn:P-S2}  & $-A^2 \left( \frac{1-\delta}{1+\delta} \right) ^2 dt^2 + \left( \frac{(1+\delta)^2}{1+\frac{r^2}{a^2}} \right) ^2 (dr^2 + r^2d\Omega ^2)$ \\
                     & where $ \delta = k\sqrt{\frac{1+\frac{r^2}{a^2}}{1+n^2\frac{r^2}{a^2}}} $  \\ \hline
D-F \cite{kn:D-F} & $-A\left((1+cr^2)^2 + B\left[ (7+3cr^2)f + 4\ln\left(\frac{(3-cr^2)f}{1+cr^2}\right) (1+cr^2)^2\right] \right)^2 dt^2 +$ \\
                  & $ + \left( 1-\frac{8}{7}cr^2\frac{3+cr^2}{(1+cr^2)^2}\right) ^{-1} dr^2 + r^2d\Omega^2 $ \\
                  & where $ f = \sqrt{7-10cr^2-c^2r^4}$ \\ \hline 
H-B I~\footnotemark \cite{kn:H-B} & $-D^2(Ar^2+B)^{-\frac{a}{1-\alpha}}\left( \frac{(Ar^2+B)^{\beta}}{4A\beta} + C \right)^2 dt^2 + $\\
		    & $ (Ar^2+B)^{\frac{b}{1-\alpha}}(dr^2 + r^2d\Omega ^2)$, \\
		    & where $\alpha = \frac{\half (b^2-a^2) -ab+b-a}{b-a}$ and $\beta = \frac{a+b}{1-\alpha} + 1$  \\ \hline
\end{tabular}
}
\end{center}

\addtocounter{footnote}{-1}\footnotetext{Eq. (38) of \cite{kn:D-P-P} has an obvious misprint in $y$.}
\addtocounter{footnote}{1}\footnotetext{As with Bayin VII (for which this is an extension) H-B I is probably contained within Kuch5 XIII.}

\begin{center}
{\noindent
\begin{tabular}{|r|l|} \hline \hline
{\bf name [ref.]} & {\bf metric} \\ \hline
H-B II \cite{kn:H-B} & $-\left( \frac{1-f}{1+f}\right) ^2 dt^2 + $\\
		     & $ (1+f)^4B^2\left[\frac{A^2}{2K}\left(-\frac{1}{2f^2} + \frac{f^2}{2} - 2\ln f\right)+C\right] ^{-2}(dr^2 + r^2d\Omega ^2)$, \\
		     & where $f = \frac{A}{\sqrt{1+Kr^2}}$\\ \hline
F-S~\footnotemark \cite{kn:F-S} & $-D^2\left( (B-A\sqrt{1+Cr^2})\cos \sqrt{1+Cr^2} + (A+B\sqrt{1+Cr^2})\sin\sqrt{1+Cr^2} \right) ^2 dt^2 + $ \\
                  & $ (1+Cr^2) dr^2 + r^2d\Omega ^2$ \\ \hline
P-P I \cite{kn:P-P} & $-\left( a+br^{\frac{2}{\sqrt{3}}}\right) dt^2 + \frac{\left( a+\frac{1+\sqrt{3}}{\sqrt{3}}br^{\frac{2}{\sqrt{3}}} \right)^{\sqrt{3}-1}}{a+br^{\frac{2}{\sqrt{3}}}} \times  $ \\
                    & $ \times \left( Ar^2 + \frac{\left( a+\frac{1+\sqrt{3}}{\sqrt{3}}br^{\frac{2}{\sqrt{3}}} \right)^{\sqrt{3}-1}\left( a(\sqrt{3}-1) - \frac{1+\sqrt{3}}{\sqrt{3}}br^{\frac{2}{\sqrt{3}}} \right)}{a^2(\sqrt{3}-1)} \right)^{-1} dr^2 + r^2d\Omega ^2$ \\ \hline
P-P II \cite{kn:P-P} & $-\left( a+br^{\frac{\sqrt{17}-1}{4}}\right) dt^2 + \left( \left(a+br^{\frac{\sqrt{17}-1}{4}} \right) \left[\frac{\sqrt{17}+5}{2a} \times \right. \right.  $ \\
                    & $ \left. \left. \times \left( 1 - \frac{\sqrt{17}+1}{4a}f+\frac{f^2}{2a^2}\right) + Ar^2f^{\frac{-\sqrt{17}+3}{2}} \right] \right)^{-1} dr^2 + r^2d\Omega ^2$, \\
                    & where $f = a + \left( \frac{\sqrt{17}+7}{8}\right) br^{\frac{\sqrt{17}-1}{4}} $ \\ \hline 
K-O III \cite{kn:K-O} & $-A(1+ar^2)^2 dt^2 + dr^2 + r^2d\Omega ^2$ \\ \hline
K-O VI \cite{kn:K-O} & $-A(d+y)^2 dt^2 + [(1+br^2)(1+Cr^2M)]^{-1} dr^2 + r^2d\Omega ^2$, \\ 
		     & where $M = \frac{(4 + \sqrt{D^2 + 8} + D + 4by)^{2\alpha}}{(2r^2+d+Dy)^{\alpha +1}}$, \\ 
		     & $D=bd-1, \alpha = \frac{D}{\sqrt{D^2+8}}$ and $ y=\frac{r^2}{1+\sqrt{1+br^2}}$ \\ \hline
K-O VII \cite{kn:K-O} & $-A(1+br^2)(D+N)^2 dt^2 + \frac{1+2br^2}{1+br^2} dr^2 + r^2d\Omega ^2$, \\ 
                      & where $N = -\frac{\sqrt{1+2br^2}}{1+br^2} + \sqrt{2}\ln (\sqrt{1+2br^2}+\sqrt{2+2br^2})$ \\ \hline 
Burl I \cite{kn:Burl} & $-A(1+r^2)^{\frac{4(a+1)}{2a^2+4a+1}} dt^2 + (1+r^2)^{\frac{4a}{2a^2+4a+1}} (dr^2 + r^2d\Omega ^2)$, \\ \hline
Burl II \cite{kn:Burl} & $-A\sqrt[3]{ \frac{(a+r^2)^2}{(a+\frac{3+\sqrt{3}}{2} + r^2)(a+\frac{3-\sqrt{3}}{2} + r^2)} } dt^2 + B \left( \frac{a+\frac{3+\sqrt{3}}{2} + r^2}{a+\frac{3-\sqrt{3}}{2} + r^2} \right) ^{\sqrt{3}} (dr^2 + r^2d\Omega ^2)$, \\ \hline
Pant I \cite{kn:Pant} & $-\left( cr^{2a+2b} + fr^{2a-2b} \right) ^{\frac{2}{m}} dt^2 + \frac{(2a+2b+m)c+(2a-2b+m)fr^{-4b}}{(c+fr^{-4b})^2} \times $ \\
                      & $ \times \left(  \frac{Ar^{-\frac{8b}{m}}}{(c+fr^{-4b})^{\frac{2}{m}}} - \frac{m^2}{4bc} \right) ^{-1} dr^2 + r^2d\Omega ^2,$ where $a=\frac{4+12m-3m^2}{8(m+2)},$ \\
                      & and $ b= -\frac{1}{8(m+2)}\sqrt{16-96m+152m^2-24m^3+m^4} $ \\ \hline
Pant II \cite{kn:Pant} & $-\left( ar^{\frac{l-1}{2}+2b}+ cr^{\frac{l-1}{2}-2b} \right) ^{2} dt^2 + \frac{\left( \half (l+1)+2b\right) a + \left( \half (l+1) -2b\right) cr^{-4b}}{Ar^{\frac{13-3l}{2}-2b} - \frac{4a}{3l-13+4b} - \frac{4c}{3l-13-4b}r^{-4b}} dr^2 + r^2d\Omega ^2,$ \\ 
                       & where $3 \leq l \leq 4,$ and $ b= -\frac{1}{4}\sqrt{l^2+2l-15}      $ \\ \hline \hline
\end{tabular}
}
\end{center}

\footnotetext{The metric of Duorah and Ray \cite{kn:D-R} did not satisfy Einstein's equations and this metric is the correction by \cite{kn:F-S}.}

\newpage

\section{Analysis} \label{sec-anal}
\bigskip

The following tables contain the metric name and state whether or not
the metric satisfies the criteria explained in the introduction\footnote{In the Table, (iff) means if and only if
the condition in the bracket is satisfied, positive def. means positive definite, and M.D. (example) means monotonically
decreasing and a specific example is given in the bracket.}. Each metric
is analyzed until a failure occurs.  There is no column for criterion four.
If a solution was found to be cosmological, a footnote states this,
and no further analysis was completed. Although not a requirement, a plus
sign in the last column is used to identify a solution where
the sound speed monotonically decreases to the boundary.  The
metrics were analyzed with the package \textit{GRTensorII} \footnote{GRTensorII is a package which runs within MapleV. It is entirely distinct from packages distributed with MapleV and must be obtained independently. The GRTensorII software and documentation is distributed freely on the World-Wide-Web from the address {\tt http://www.astro.queensu.ca/\~{}grtensor/GRHome.html} 
or {\tt http://www.maths.soton.ac.uk/\~{}dp/grtensor/}.  Worksheets which demonstrate some calculations reported here, and the associated metric files,  can be downloaded from these sites.} \cite{kn:GRT}
in conjunction with MapleV \cite{kn:Maple}.

\small
\begin{center}
{\noindent
\begin{tabular}{|r|c|c|c|c|c|} \hline \hline
{\bf name} & {\boldmath $G^{r}_{r} = G^{\theta}_{\theta}$} {\bf (iff)} & {\bf Regular at}
&{\boldmath $p$} {\bf and} {\boldmath $\rho$} & {\boldmath $p$} {\bf and} {\boldmath $\rho$} {\bf M.D.} & {\boldmath $\frac{dp}{d\rho} < 1$} \\
& &{\bf the origin (iff)}& {\bf positive def.} & {\bf (example)} & \\ \hline
Schw Int.       & Y & Y & Y & Y ($\{A,B,R\}$ & N\\ 
                &   &   &  & $= \{1,\half,2\}$) & \\ \hline
Einstein        & Y & Y & N & &\\ \hline
de Sitter       & Y & Y & N & &\\ \hline
Kottler         & Y & Y $(m=0)$ & N & &\\ \hline 
Tolman IV       & Y & Y & Y  & Y ($\{A,R\}$ & Y \\  
		&   &   &             & $= \{5,10\}$) & \\ \hline
Tolman V        & Y & Y $(n=0)$ & N & &\\ \hline 
Tolman VI       & Y & Y $(n = \pm 1)$ & N & &\\ \hline 
Tolman VII      & Y & Y & Y & Y ($\{A,R,B,C\}$ & Y+ \\  
		&   &   &   & $= \{1,0.54,1,20\}$) & \\ \hline
Tolman VIII     & Y  & Y ($\{a,b,m\}=$   & & &\\  
		&    & $\{2,0,0\}$ & & &\\ 
		&    & $\rightarrow$ Tolman II) & & &\\ \hline 
N-P-V Ia        & Y & Y   &   & &\\ 
		&   & ($n=2,B=1 \rightarrow $ & & & \\
                &   & Sch. Int.) & & & \\ \hline
N-P-V Ib        & Y  & N  &   & &\\ \hline
N-P-V Ic        & N &  &  &  & \\ \hline 
N-P-V IIa       & Y & Y    &   & &\\ 
		&   & ($k=0 \rightarrow $ & & & \\
		&   & flat space) & & & \\ \hline
\end{tabular}
}
\end{center}

\begin{center}
{\noindent
\begin{tabular}{|r|c|c|c|c|c|} \hline \hline
{\bf name} & {\boldmath $G^{r}_{r} = G^{\theta}_{\theta}$} {\bf (iff)} & {\bf Regular at}
&{\boldmath $p$} {\bf and} {\boldmath $\rho$} & {\boldmath $p$} {\bf and} {\boldmath $\rho$} {\bf M.D.} & {\boldmath $\frac{dp}{d\rho} < 1$} \\
& &{\bf the origin (iff)}& {\bf positive def.} & {\bf (example)} & \\ \hline
N-P-V IIb       & Y & N  &   & &\\ \hline
N-P-V IIc       & Y & N &  &  & \\ \hline
P-V IIa        & Y & Y  & Y  & Y ($\{A,B,b,c,d\}$ & Y+ \\
	       &   &    &    & $=\{1,3,2,1,1\}$)  & \\ \hline
P-V IIb        & Y & Y  & N &  &\\ \hline

P-V IIc        & Y ($c=0 $ & Y  & Y & N & \\
	       & or $b=0 \rightarrow$  &  &   & & \\ 
               & Sch. Int.  &  &   & & \\ \hline 
P-V IV         & Y & N  &   & &\\ \hline
P-V V          & N &    &   & &\\ \hline 
Wyman I         & Y $(N=2 \rightarrow $ &       & & &\\
                    & Sch. Int.)  & &  & &\\ \hline
Wyman IIa       & Y & Y $(n = 1)$     & Y & Y ($\{a,A,B\} $ & Y \\
		&   &                 &   & $= \{-1,1,-\half\}$) & \\ \cline{3-6} 
		&   & Y $(n=-1)$      & Y & Y ($\{a,A,B\} $ & Y \\ 
		&   &                 &   & $= \{-1,\half,-1\}$) & \\ \hline
Wyman IIb       & Y $(a=0)$ & N &  & &\\ \hline 
Wyman IIc       & N &  &  & &\\ \hline 
Wyman III       & Y $(n=1)$ & N &  & &\\ \hline 
Wyman IVa       & Y & N & & &\\ \hline 
Wyman IVb       & Y $(A=0)$ & N & & &\\ \hline
Nariai III      & Y $(k=1 \rightarrow$ &   &  & & \\ 
                & Sch. Int.)          &   &  & & \\ \hline
Nariai IV       & Y & Y & Y & Y ($\{a,b,M\} = $ & Y+\\ 
		&   & $(A = \cos ^{2}(b))$  &  &$\{1,2.4,1\}$) & \\ \hline
Nariai VI       & Y & Y ($\alpha = 0 \rightarrow$ &  &  & \\ 
                &   & Sch. Int.)  &  &  & \\ \hline
Nariai VII      & Y & N &  &  & \\ \hline
Nariai VIII     & Y & N &  &  & \\ \hline
Nariai IX       & Y & N &  &  & \\ \hline
Nariai X        & Y & N &  &  & \\ \hline
Buch1          & Y & Y & Y & Y ($\{A,B,C\} =$ & Y \\
	       &   &   &   & $\{1,\half,1\}$) & \\ \hline
Buch2          & Y & Y ($a=0 \rightarrow $  &  & &\\
	       &   & flat space) & & & \\ \hline
Mehra           & Y & Y & Y & Y ($\{a,\rho_c\} =$ & Y+ \footnotemark \\ 
		&   &   &   & $\{0.99,0.1\}$) & \\ \hline
Buch3         & Y  & Y\footnotemark & N & &\\ \hline
 \end{tabular}
}
\end{center}
\addtocounter{footnote}{-1}\footnotetext{Mehra's solution is the only one where the pressure, density and speed of sound all equal zero at the boundary $r_b = a$.}
\addtocounter{footnote}{1}\footnotetext{Buch3 is regular iff $\frac{c}{a} = \frac{b-\sqrt{b^2-4ac}}{b+\sqrt{b^2-4ac}}$.}

\begin{center}
{\noindent
\begin{tabular}{|r|c|c|c|c|c|} \hline \hline
{\bf name} & {\boldmath $G^{r}_{r} = G^{\theta}_{\theta}$} {\bf (iff)} & {\bf Regular at}
&{\boldmath $p$} {\bf and} {\boldmath $\rho$} & {\boldmath $p$} {\bf and} {\boldmath $\rho$} {\bf M.D.} & {\boldmath $\frac{dp}{d\rho} < 1$} \\
& &{\bf the origin (iff)}& {\bf positive def.} & {\bf (example)} & \\ \hline
Kucha    & Y $(C=0)$ & N  &  & &\\ 
	 &   &    &  & &\\ \hline 
Kuchb Ia    & Y & Y $(B=1 \rightarrow$ &  & &\\ 
            &   &  Sch. Int.) & & &\\ \hline
Kuchb Ib    & Y $(B=0)$ & N &  & &\\ \hline
Kuchb Ic    & N &  &  & &\\ \hline 
Kuch1 Ib    & Y $(C=2)$ & N &  & &\\ \hline 
Kuch1 Id    & Y & N &  & &\\ \hline
Kuch1 IIIb  & N &  &  & &\\ \hline
Kuch1 IV    & N &  &  & &\\ \hline
Kuch1 V     & N &  &  & &\\ \hline
Kuch2 I    & Y & Y($A=0 \rightarrow$ &  &  & \\
           &   &  Tolman I) & & & \\ \hline
Kuch2 III   & Y  & Y & Y  & Y ($\{A,C\} = $ & Y+ \\ 
	    &    &   &   &  $ \{5,-3\}$) & \\ \hline
Kuch2 IV   & N &  &  &  & \\ \hline 
Kuch2 VI   & Y & N &  &  & \\ \hline 
Kuch2 VII  & Y & N &  &  & \\ \hline 
Whittaker       & Y & Y & Y & N  &\\ \hline
B-L             & Y & N & & &\\ \hline
Kuch68 I       & Y & N & & &\\ \hline
Kuch68 II      & Y & N & & &\\ \hline
Leib I          & Y $(a=0 \rightarrow $ &  &  &  & \\
		& Toman III)            &  &  &  & \\ \hline
Leib IV         & Y ($a=0 \rightarrow$ &  &  &  & \\ 
		& Tolman I) &  &  &  & \\ \hline
Heint IIa       & Y & Y & Y & Y ($\{a,A,C\}$ & Y \\ 
		&   &   &   & $= \{1,1,1\}$) & \\ \hline
Heint IIb  & Y $(a=0 \rightarrow$       & Y  & N & &\\
	   & flat space  &    &   & & \\
	   & or $C=0$) & & & & \\ \hline
Heint IIIa & Y $(b=0 \rightarrow$ &  &  &  & \\ 
	   & flat space)          &  &  &  & \\ \hline
Heint IIIb & Y $(a=C=0)$ & N &  & &\\ \hline 
Heint IIIe & Y & Y & N & &\\ \hline 
Kuch3 Ia   & N &  &  &  & \\ \hline
Kuch3 Ib   & Y & N &  &  & \\ \hline 
Kuch3 Ic   & N &  &  &  & \\ \hline 
Kuch3 II   & Y $(B=0)$ & N &  &  & \\ \hline 
Kuch3 III  & N &  &  &  & \\ \hline 
\end{tabular}
}
\end{center}

\begin{center}
{\noindent
\begin{tabular}{|r|c|c|c|c|c|} \hline \hline
{\bf name} & {\boldmath $G^{r}_{r} = G^{\theta}_{\theta}$} {\bf (iff)} & {\bf Regular at}
&{\boldmath $p$} {\bf and} {\boldmath $\rho$} & {\boldmath $p$} {\bf and} {\boldmath $\rho$} {\bf M.D.} & {\boldmath $\frac{dp}{d\rho} < 1$} \\
& &{\bf the origin (iff)}& {\bf positive def.} & {\bf (example)} & \\ \hline
Kuch3 IV   & N &  &  &  & \\ \hline 
Kuch3 V    & N &  &  &  & \\ \hline 
Kuch71 I   & N &  &  &  & \\ \hline 
Kuch71 II  & Y & Y & Y & N & \\ \hline 
Kuch71b    & Y & N &   &   & \\ \hline 
Kuch5 I    & N &  &  &  & \\ \hline
Kuch5 IV   & Y $(A=0 \rightarrow $ &  &  &  & \\
           & Sch. Int.            &  &  &  &  \\ \hline
Kuch5 VIII & Y $(a=0 \rightarrow $ &  & &  & \\
	   & flat space) & & & & \\ \hline
Kuch5 IX      & Y & N &  &  & \\ \hline
Kuch5 XI      & N &  &  &  & \\ \hline
Kuch5 XII     & N &  &  &  & \\ \hline
Kuch5 XIII    & Y & Y & Y\footnotemark &  & \\ 
              &   & ($D = |B|^{\frac{2(a-1)}{2a^2-1}}$)  & &  & \\ \hline
Kuch5 XV      & Y $(B=0 \rightarrow $ &  &  &  & \\
              & flat space) & & & & \\ \hline
Kuch5 XVI     & N &  &  &  & \\ \hline
R-R I         & Y & Y & N\footnotemark & &\\ \hline 
R-R II        & Y & N &  & &\\ \hline 
R-R III       & Y & Y ($A = B \pm 1$)  & N & &\\ \hline 
R-R IV        & Y & Y $(B=1)$ & N & &\\ \hline 
R-R V         & Y & N &  & &\\ \hline 
R-R VI        & Y & Y ($n=1,A=1$ & N  & &\\ 
              &   & or $n=-1,B=1$) & & &\\ \hline 
R-R VII       & Y & Y\footnotemark  & N & &\\ \hline
R-R VIII      & Y & Y $(\{a,b,m,B\}=$ & N  & &\\ 
              &   &  $\{2,0,0,1\})$ &  & &\\ \hline
Kuch73 I      & Y & N &  &  & \\ \hline 
Kuch73 II     & N &   &  &  & \\ \hline 
K-N-B         & Y & Y & Y & N & \\ \hline 
G-G           & Y & Y\footnotemark & Y & N & \\ \hline
\end{tabular}
}
\end{center}

\addtocounter{footnote}{-3}\footnotetext{Kuch5 XIII is a cosmological solution.}
\addtocounter{footnote}{1}\footnotetext{All these solutions have either a negative
pressure or density, as was expected since Roy-Rao used the Buchdahl Transformation
(BT) \cite{kn:Buch54,kn:Buch56}.  A generalization of the BT that allows for positive pressure
and density was completed by Stewart \cite{kn:Stew}.}
\addtocounter{footnote}{1}\footnotetext{Regular iff $B\sin \ln \left(\half \frac{4R^2-A^2}{R^2C}\right) =1$.}
\addtocounter{footnote}{1}\footnotetext{G-G is regular iff $\frac{B}{b^2(2+d)}\left(\frac{-b^2d-1-c}{-b^2d-1+c}\right)^{(-b^2d-2b+1)/c} =1$.}

\begin{center}
{\noindent
\begin{tabular}{|r|c|c|c|c|c|} \hline \hline
{\bf name} & {\boldmath $G^{r}_{r} = G^{\theta}_{\theta}$} {\bf (iff)} & {\bf Regular at}
&{\boldmath $p$} {\bf and} {\boldmath $\rho$} & {\boldmath $p$} {\bf and} {\boldmath $\rho$} {\bf M.D.} & {\boldmath $\frac{dp}{d\rho} < 1$} \\
& &{\bf the origin (iff)}& {\bf positive def.} & {\bf (example)} & \\ \hline
Bayin III     & N &  &  & &\\ \hline 
Bayin VI      & Y & Y ($BC^{b} = 1)$ & Y & N &\\ \hline 
Gold I        & Y & Y\footnotemark & Y & N & \\ \hline
Gold II       & Y & Y\footnotemark & Y & N & \\ \hline
Gold III        & Y & Y & Y & Y ($\{a,b,A\}$ & Y+ \\
                &   & ($B = \frac{(e^{2a}+1)^2}{(e^a+1)^4}$) & & $= \{ 2,1,1\}$) & \\ \hline
M-W I           & Y & Y  & Y  & Y ($\{a,A,B,R\}$  & Y+ \\
                &   &    &    & $= \{ 1,-1,-1,3\}$) &    \\ \hline
M-W II          & Y $(A=0 \rightarrow$  &   &   &   &\\
                & Tolman IV)            &   &   &   & \\  \hline
M-W III         & N &  &   &   &\\ \hline
K-K             & N &  &   &   &\\ \hline
Stewart         & Y & Y & Y & N & \\ \hline
P-S Ia          & Y & Y ($a=-2$ &  & &\\ 
		&  & $\rightarrow$ Tolman IV) &  & &\\ \hline
P-S Ib          & Y & N &  & &\\ 
		&   &   &  & &\\ \hline
Durg IV         & Y & Y & Y & Y ($\{C,A,K\} =$        & Y+ \\
		&   &   &   & $\{1,1,-\frac{1}{4}\}$) & \\ \hline
Durg V          & Y & Y & Y & Y ($\{C,A,K\} =$        & Y+ \\
                &   &   &   & $\{1,1,-\half\}$)       & \\ \hline
Whitman II       & Y & Y & Y & Y ($\{a,R,A,B\} =$      & Y \\
                &   &   &   & $\{4,1,1,1 \}$) &   \\ \hline
Whitman III    & N &  &  & &\\ \hline
Whitman IV     & Y & Y & Y & Y ($\{a,R,b,C\} =$      & Y \\
               &   &   &   & $\{1,4,-1,\frac{1}{4}\}$) &   \\ \hline
D-P-P I         & Y & Y  & Y & N & \\ \hline
V-P II          & Y ($A=0 \rightarrow$ &  &  & &\\
                & flat sp) & & & & \\ \hline
P-S2            & Y  & N  &   & & \\ \hline
D-F             & Y $(B=0)$ & Y & Y & Y $(\{ A,B,c\} =$ & N \\
                &           &   &   & $\{ 1,0,1\} )$   &    \\ \hline
H-B I           & Y & Y & N & &\\
		&   & $(B^{b/(1-\alpha )}=1)$ & & & \\ \hline
\end{tabular}
}
\end{center}

\addtocounter{footnote}{-1}\footnotetext{Gold I is regular iff $\frac{B}{a+2c^2+2ac}\left(\frac{1+ac-bc}{1+ac+bc}\right)^{-(a+1)/b} =1$.}
\addtocounter{footnote}{1}\footnotetext{Gold II is regular iff $B e^{(-2-\sqrt{2}c)a/(c^2-1)} \left(\frac{2e^a}{e^{2a}(1+c)+1-c}\right) ^{(-2-\sqrt{2}c)/(c^2-1)} =1$.}

\begin{center}
{\noindent
\begin{tabular}{|r|c|c|c|c|c|} \hline \hline
{\bf name} & {\boldmath $G^{r}_{r} = G^{\theta}_{\theta}$} {\bf (iff)} & {\bf Regular at}
&{\boldmath $p$} {\bf and} {\boldmath $\rho$} & {\boldmath $p$} {\bf and} {\boldmath $\rho$} {\bf M.D.} & {\boldmath $\frac{dp}{d\rho} < 1$} \\
& &{\bf the origin (iff)}& {\bf positive def.} & {\bf (example)} & \\ \hline
H-B II          & Y & Y\footnotemark & N & &\\ \hline 
F-S             & Y & Y & Y & Y ($\{A,B,c,D \} = $ & Y \\ 
                &   &   &   & $\{1,\frac{1}{2},\frac{1}{2},1 \}$) &   \\ \hline
P-P I           & Y & Y & Y ($\{ a,b\} = \{ 1,0\} $ & &\\
                &   &   & $\rightarrow$ Einstein    & & \\ \hline
P-P II          & Y & Y $(b=0)$     & N  & &\\
                &   &               &    & & \\ \hline
K-O III         & Y & Y & N & &\\ \hline
K-O VI          & Y & Y & N & &\\ \hline 
K-O VII         & Y ($b=0 \rightarrow$ &  &  & & \\ 
                & flat space) & & & & \\ \hline
Burl I          & N &  &   &   &   \\ \hline
Burl II         & N &  &   &   &   \\ \hline
Pant I          & Y & Y $(m=2 \rightarrow$ &   &  & \\
                &   & Tolman IV            &   &  & \\ \hline
Pant II         & Y & N  &   & &\\ \hline
\end{tabular}
}
\end{center}

\footnotetext{H-B II if regular iff $\frac{16B^2K^2(A+1)^4}{A^4+4CK-4A^2\ln A -1}=1$.}

\normalsize
\newpage

\section{Discussion} \label{sec-disc}
\bigskip

Of the 127 solutions studied, only 16 satisfy all the criteria set out in section I, 
and 9 of these 16 were found to have the additional property of having 
the sound speed monotonically decrease with radius. About one solution in five
failed to give isotropic pressure\footnote{Since the Einstein tensor is diagonal for
the metrics under consideration, \textit{any} such metric can be considered as an anisotropic ``solution".}, and of those that did one quarter failed to be
regular at the origin. Of the solutions that were regular at the origin one quarter
failed to give positive energy density and pressure there. Of the solutions that formally
allow an integration from the origin about one in five failed to produce monotone
decreasing functions $\rho(r)$ and $p(r)$.

It is fair to say then that most of the spherically symmetric perfect fluid ``exact solutions" of
Einstein's field equations that are in the literature are of no physical interest. The reason
for this can be traced to the  Tolman-Oppenheimer-Volkoff equation (\ref{eqn:TOV}). Whereas it is now
known that for a given value of $0 < p(0) < \infty$ there exists a unique global solution to
equation (\ref{eqn:TOV}) \cite{kn:rs}, this does \textit{not} mean that closed-form solutions are easy to find.
Indeed ``one soon has to use numerical methods" \cite{kn:ESEFE}. For the most part the spacetimes
examined in this paper were obtained by various simplifying assumptions developed in order to obtain
an ``exact solution". What we have shown here is that this procedure almost never leads to a physically 
interesting conclusion.

The content of this work would make a useful addition to a database of solutions of Einstein's field
equations. The pioneering effort in this field is the ``On-Line Invariant Classification Database" developed
by Jim Skea. \footnote{ The main site in Brazil is at {\tt edradour.symbcomp.uerj.br/}, 
and mirrors are at {\tt www.astro.queensu.ca/~jimsk/}  and {\tt www.maths.soton.ac.uk/\~{}rdi/database}. A keyword search
with ``static sphere" returns 7 Petrov type $D$ and 2 Petrov type $O$ spacetimes. A search with ``perfect fluid"
returns 31 type $D$ and 10 type $O$ spacetimes.}

\section*{Acknowledgments}

We would like to thank M. MacCallum for his comments, suggestions and corrections,
J. Skea for a preprint (\cite{kn:FS} with M. Finch) which contained a number of references
we had not found, and H. Knutsen, A. Krasinski and H. Stephani for comments. This work was supported in part by grants (to KL) from the Natural 
Sciences and Engineering Research Council of Canada and the Advisory Research Committee 
of Queen's University. 

\newpage
\section*{Appendix A}
\bigskip

Consider the general diagonal, static, spherically symmetric metric \footnotemark:
\footnotetext{Any static spherically symmetric line element can be brought into this form \cite{kn:takeno}.}
\be
ds^2 = -A(r) dt^2 + B(r) dr^2 + C(r)r^2d\Omega ^2. \nonumber
\ee
The regularity conditions are
\be
A(0) = \mbox{const.}, B(0) = C(0) = 1, \nonumber
\ee
and
\be
A'(0) = B'(0) = C'(0) = 0. \nonumber
\ee
Note that the metric (1), if regular, is conformally flat at the origin
so that the only non-vanishing invariants there are the Ricci invariants. In
terms of the trace-free Ricci tensor $S^{a}_{b}$ defined by
\be
S^{a}_{b} = R^{a}_{b}-\delta^{a}_{b}R/4,
\ee
where $R^{a}_{b}$ is the Ricci tensor, and $R$ the Ricci scalar,
the Ricci invariants at the origin reduce to
\be
R=3(B^{''}(0)-A^{''}(0)-3C^{''}(0)),\nonumber
\ee
\be
r_{1} \equiv S^{b}_{a}S^{a}_{b}/4 = 3(B^{''}(0)+A^{''}(0)-3C^{''}(0))^{2}/2^{4},\nonumber
\ee
\be
r_{2} \equiv -S^{b}_{a}S^{c}_{b}S^{a}_{c}/8 = 3(B^{''}(0)+A^{''}(0)-3C^{''}(0))^{3}/2^{6},\nonumber
\ee
\be
r_{3} \equiv S^{b}_{a}S^{c}_{b}S^{d}_{c}S^{a}_{d} = 21(B^{''}(0)+A^{''}(0)-3C^{''}(0))^{4}/2^{10}.
\ee
The energy density and pressure at the origin reduce to
\be
8\pi \rho(0) = 3B^{''}(0)/2-9C^{''}(0)/2,
\ee
and
\be
8\pi p(0)=A^{''}(0)-B^{''}(0)/2+3C^{''}(0)/2
\ee
respectively. In \textbf{isotropic coordinates} $ C(r) = B(r)$, and $C(r) = 1$ in \textbf{curvature coordinates}.

\section*{Appendix B}
\bigskip

The equilibrium is governed by the well known Tolman-Oppenheimer-Volkoff equation
\be
\frac{dp(r)}{dr} = -\frac{(p(r)+\rho(r) )(m(r)+4\pi p(r) r^3)}{r(r-2m(r))}  \label{eqn:TOV}
\ee
where $m(r)$ is the effective gravitational mass
\be
m(r)=4\pi\int ^{r}_{0} \rho(s)s^2ds.
\ee
It follows that
\be
dp/dr| _{r=0} = 0,
\ee
and that
\be
d^{2}p/dr^{2}| _{r=0} =-4\pi(\rho(0)+3p(0))(\rho(0)+p(0))/3 ,
\ee
so that the pressure is locally maximal with respect to $r$ at the origin. If we assume that there is a smooth ($C^{2}$)
equation of state $p(\rho)$ then we have
\be
dp/dr=(dp/d\rho)(d\rho/dr),
\ee
and
\be
d^{2}p/dr^{2}=(d^{2}p/d\rho^{2})(d\rho/dr)^{2}+(dp/d\rho)(d^{2}\rho/dr^{2}).
\ee
As a result, $\rho$ is locally maximal with respect to $r$ at $r=0$ or (not and)
$p$ is locally maximal with respect to $\rho$ at $r=0$.
\par
If
\be
dp/d\rho > 0\label{eqn:in}
\ee
for $p>0$, it is known that for a given value of $0 < p(0) < \infty$ there exists a unique global solution to
equation (11) \cite{kn:rs}. More recently this has been established without condition (17)
\cite{kn:br}.
Without an equation of state, specification of $\rho(r)$ (or $m(r)$) reduces the problem to a single differential
equation (11). The resultant implied equation of state is almost always of no interest.

\section*{Appendix C}
\bigskip

The following table contains the original solution in the first
column.  The second column contains rediscoveries where the authors were unaware of the 
previous work.  For example,
Kuchowicz \cite{kn:Kuch75} rederived Wyman IIa \cite{kn:Wyman}, but
gave credit to Adler \cite{kn:Adler} as the originator, while
Leibovitz \cite{kn:Leib} also rederived Wyman IIa but was unaware of Wyman's work. We do not claim that 
this table is complete.

\vspace{0.25in}

\begin{tabular}{|l|l|} \hline \hline
\bf {original} & \bf {rederived by} \\ \hline
Tolman IV \cite{kn:Tolman} & K-GT-P \cite{kn:K-GT-P} \\ \hline
Tolman V \cite{kn:Tolman} & D-G \cite{kn:D-G} \\ \hline
Tolman VI \cite{kn:Tolman}&P-V \cite{kn:P-V}, Klein \cite{kn:Klein} \\ \hline
Tolman VII \cite{kn:Tolman} & D-P-P II \cite{kn:D-P-P} \\ \hline
Wyman IIa $(n=1)$ \cite{kn:Wyman} & Leib II \cite{kn:Leib}, Heint IIId \cite{kn:Heint}, DurgII \cite{kn:Durg}, \\
                                  & Adler \cite{kn:Adler}, A-C \cite{kn:A-C}, Kuch75 \cite{kn:Kuch75},  \\
                                  & Whitman \cite{kn:Whitman} \\ \hline
Buch1 \cite{kn:Buch1} & D-B \cite{kn:D-B} \\ \hline
Kuch2 I \cite{kn:Kuch2} & Krori \cite{kn:Krori}, I-S II \cite{kn:I-S} \\ \hline
Kuch2 VI \cite{kn:Kuch2} &  Heint IIIc \cite{kn:Heint} \\ \hline
Kuch2 III \cite{kn:Kuch2} & Leib~\footnotemark III \cite{kn:Leib} \\ \hline
Leib I \cite{kn:Leib} & Heint I \cite{kn:Heint}  \\ \hline
Heint IIa \cite{kn:Heint} & Korkina \cite{kn:Kork}, Durg III \cite{kn:Durg} \\ \hline
Heint IIIe \cite{kn:Heint} & Bayin IV \cite{kn:Bayin} \\ \hline
Kuch5 XIII \cite{kn:Kuch5} & Bayin VII \cite{kn:Bayin} \\ \hline
K-N-B \cite{kn:K-N-B} & Bayin II \cite{kn:Bayin}  \\ \hline
\end{tabular}
\footnotetext{The metric given by Leibovitz is incorrect, although the correct form was already given by Kuchowicz.}

\end{document}